\documentclass[aps,prb,twocolumn,superscriptaddress,showpacs]{revtex4}
\usepackage{amssymb}
\usepackage{graphicx}
\usepackage{hyperref}
\usepackage{bm}
\begin{document}
\title{Single Crystal Growth, Crystallography, Magnetic Susceptibility, Heat
Capacity, and Thermal Expansion of the Antiferromagnetic $\mathbf
{S = 1}$ Chain Compound ${\rm{\mathbf{CaV_2O_4}}}$}

\author{A. Niazi}
\altaffiliation[Present address: ]{Department of Physics, Jamia Millia Islamia, New Delhi - 110025, India.}
\affiliation{Ames Laboratory, Ames, Iowa 50011, USA}
\author{S. L. Bud'ko}
\affiliation{Ames Laboratory, Ames, Iowa 50011, USA}
\affiliation{Department of Physics and Astronomy, Iowa State University,
Ames, Iowa 50011, USA}
\author{D.~L. Schlagel}
\author{J.~Q. Yan}
\author{T.~A. Lograsso}
\affiliation{Materials and Engineering Physics Program, Ames Laboratory,
Ames, Iowa 50011, USA}
\author{A.~Kreyssig}
\author{S.~Das}
\author{S. Nandi}
\author{A. I. Goldman}
\affiliation{Ames Laboratory, Ames, Iowa 50011, USA}
\affiliation{Department of Physics and Astronomy, Iowa State University,
Ames, Iowa 50011, USA}
\author{A. Honecker}
\affiliation{Institut f\"ur Theoretische Physik, Universit\"at
G\"ottingen, D-37077 G\"ottingen, Germany}
\author{R.~W. McCallum}
\affiliation{Materials and Engineering Physics Program, Ames Laboratory,
Ames, Iowa 50011, USA}
\author{M. Reehuis}
\affiliation{Helmholtz-Zentrum Berlin f\"{u}r Materialien und Energie (HZB), 
Glienicker Stra\ss e 100, 14109 Berlin, Germany}
\affiliation{Max-Planck-Institut f\"{u}r Festk\"{o}rperforschung,
Heisenbergstr.~1, D-70569 Stuttgart, Germany}
\author{O. Pieper}
\author{B. Lake}
\affiliation{Hahn-Meitner-Institut, Glienicker Str.~100, D-14109 Berlin,
Germany}
\affiliation{Institut f\"{u}r Festk\"{o}rperphysik, Technische
Universit\"{a}t Berlin, Hardenbergstr.~36, D-10623 Berlin, Germany}
\author{D.~C. Johnston}
\affiliation{Ames Laboratory, Ames, Iowa 50011, USA}
\affiliation{Department of Physics and Astronomy, Iowa State University,
Ames, Iowa 50011, USA}
\date{\today}

\begin{abstract} The compound CaV$_2$O$_4$ contains V$^{+3}$ cations with
spin $S = 1$ and has an orthorhombic structure at room temperature containing 
zigzag chains of V atoms running along the $c$-axis.   We have grown single
crystals of CaV$_2$O$_4$ and report crystallography, static magnetization,
magnetic susceptibility $\chi$, ac magnetic susceptibility, heat capacity
$C_{\rm p}$, and thermal expansion measurements in the temperature $T$ range of
1.8--350~K on the single crystals and on polycrystalline samples.  An 
orthorhombic to monoclinic structural distortion and a
long-range antiferromagnetic (AF) transition were found at
sample-dependent temperatures $T_{\rm S} \approx 108$--145~K and
$T_{\rm N}\approx 51$--76~K, respectively.  In two annealed single crystals,
another transition was found at $\approx 200$~K\@.  In one of the crystals, this transition is mostly due to V$_2$O$_3$ impurity phase that grows coherently in the crystals during annealing.  However, in the other crystal the origin of this transition at 200~K is unknown.  The $\chi(T)$ shows a broad maximum
at $\approx 300$~K associated with short-range AF ordering and the anisotropy of
$\chi$ above $T_{\rm N}$ is small.  The anisotropic $\chi(T \to 0)$ data
below $T_{\rm N}$ show that the (average) easy axis of the AF magnetic structure
is the $b$-axis.  The $C_{\rm p}(T)$ data indicate strong short-range AF
ordering above $T_{\rm N}$, consistent with the $\chi(T)$ data.  We fitted our
$\chi$ data by a $J_1$-$J_2$ $S = 1$ Heisenberg chain model, where $J_1$($J_2$)
is the (next)-nearest-neighbor exchange interaction.  We find $J_1
\approx 230$~K, and surprisingly, $J_2/J_1 \approx 0$ (or $J_1/J_2 \approx 0$).  The interaction $J_\perp$ between these $S = 1$ chains leading to long-range AF ordering at $T_{\rm N}$ is estimated to be $J_\perp/J_1 \gtrsim 0.04$.

\end{abstract}


\maketitle

\section{Introduction} 

Low-dimensional frustrated spin systems have rich phase diagrams arising
from a complex interplay of thermal and quantum fluctuations and
competing magnetic interactions at low temperatures.  While spin $S=1/2$
antiferromagnetic (AF) chains\cite{Johnston2000A} and odd-leg
ladders\cite{Johnston2000,Dagotto2001} have gapless magnetic excitations,
$S=1$ chains and $S=1/2$ even-leg ladders with nearest-neighbor (NN, $J_1$)
interactions exhibit a finite energy gap between the ground state and the
lowest excited magnetic states.  However, numerical calculations have shown
that the influence of frustrating next-nearest-neighbor (NNN, $J_2$)
interactions play a significant role and depending on the $J_2/J_1$ ratio, can
lead to incommensurate helical spin structures which may be gapped or gapless.
\cite{kaburagi, hikihara1, hikihara2, kolezhuk,mikeska}  Such a system is
described by the $XXZ$ Hamiltonian \cite{kaburagi}
\begin{equation} 
{\cal H} = \sum_{\rho=1}^2 J_{\rho}\sum_l(S^{x}_lS^{x}_{l+\rho} +
S^{y}_lS^{y}_{l+\rho} + \lambda S^{z}_lS^{z}_{l+\rho}),
\label{EqH1}
\end{equation}  
where ${\bf S}_l$ is the spin operator at the $l$th site,
$J_{\rho}$ is the AF interaction between the NN ($\rho = 1$) and NNN 
($\rho = 2$) spin pairs, and $\lambda$ is the
exchange anisotropy. For $j \equiv J_2/J_1 > 1/4$, the classical AF chain
exhibits incommensurate helical long-range ordering described by the wave
vector $q = \arccos[-1/(4j)]$ and a finite {\em vector chirality}
$\vec{\bf \kappa} = {\bf S}_i \times {\bf S}_{i+1}$ which describes the sense
of rotation (left or right handed) of the spins along the helix.  In the
large-$j$, small-$\lambda$ limit of the $S = 1$ chain, one finds a
corresponding phase\cite{hikihara1} where spin correlations decay, as required
for a one-dimensional system, although only algebraically, but chirality is
still long-range ordered.  This phase is called the \emph{chiral gapless phase}
and is seen to exist for all spin quantum numbers $S$.\cite{hikihara2,
kolezhuk}  For smaller $j$, a \emph{chiral gapped phase} is observed in the $S
= 1$ chain,\cite{hikihara1} with chiral long-range order and exponentially
decaying spin correlations.

The above chiral phases are ground state phases of a spin system.  In a
related prediction, Villain suggested three decades ago that a long-range
ordered vector chiral phase can exist above the N\'eel temperature $T_{\rm N}$
of a quasi-one-dimensional spin chain system showing helical magnetic ordering
below $T_{\rm N}$.\cite{Villain1977}  This chiral phase would have a transition
temperature $T_0 > T_{\rm N}$ that could be detected using heat capacity
measurements.\cite{Villain1977}
 
\begin{figure}[t]
\includegraphics[width=3in]{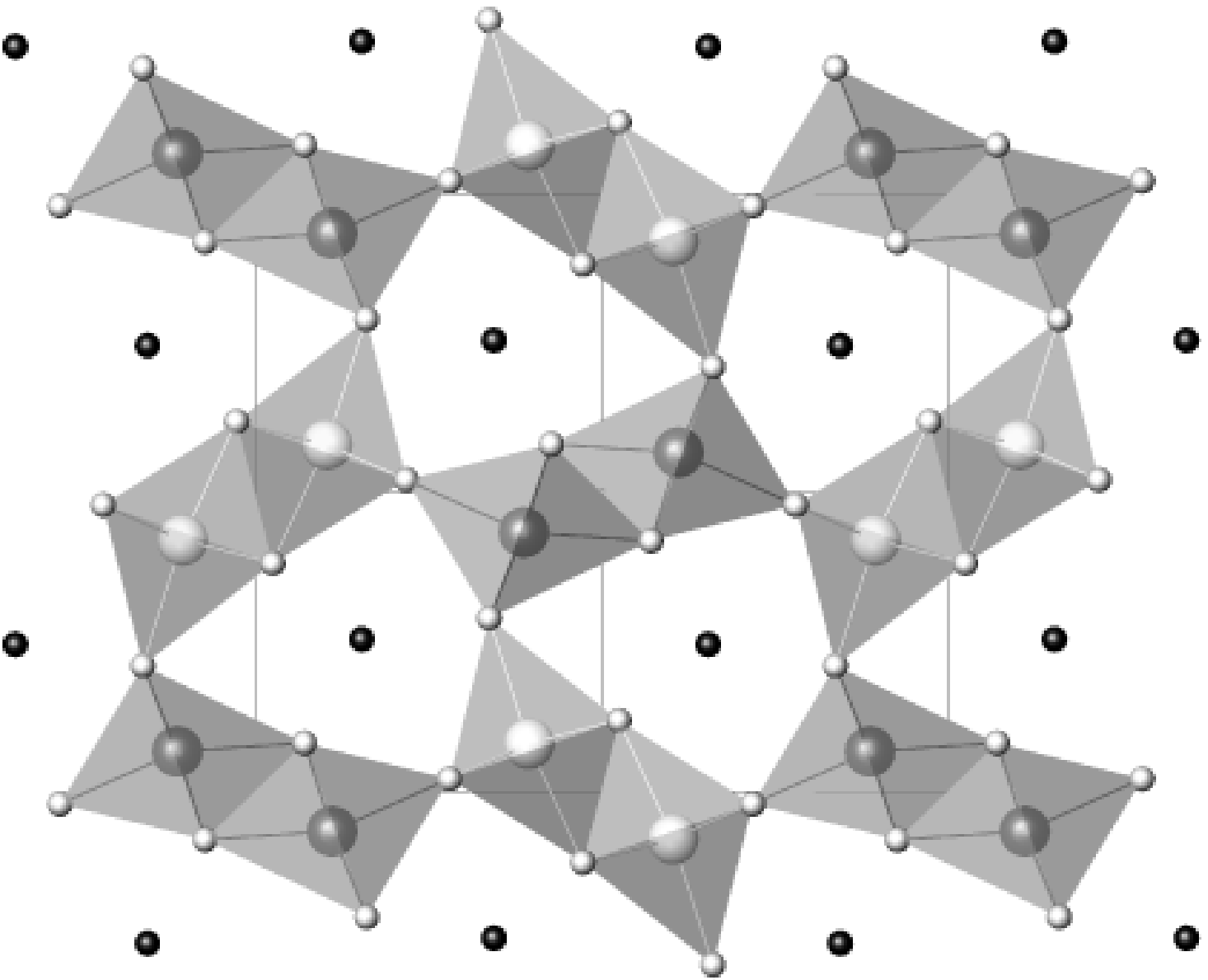}\\
\includegraphics[width=3in]{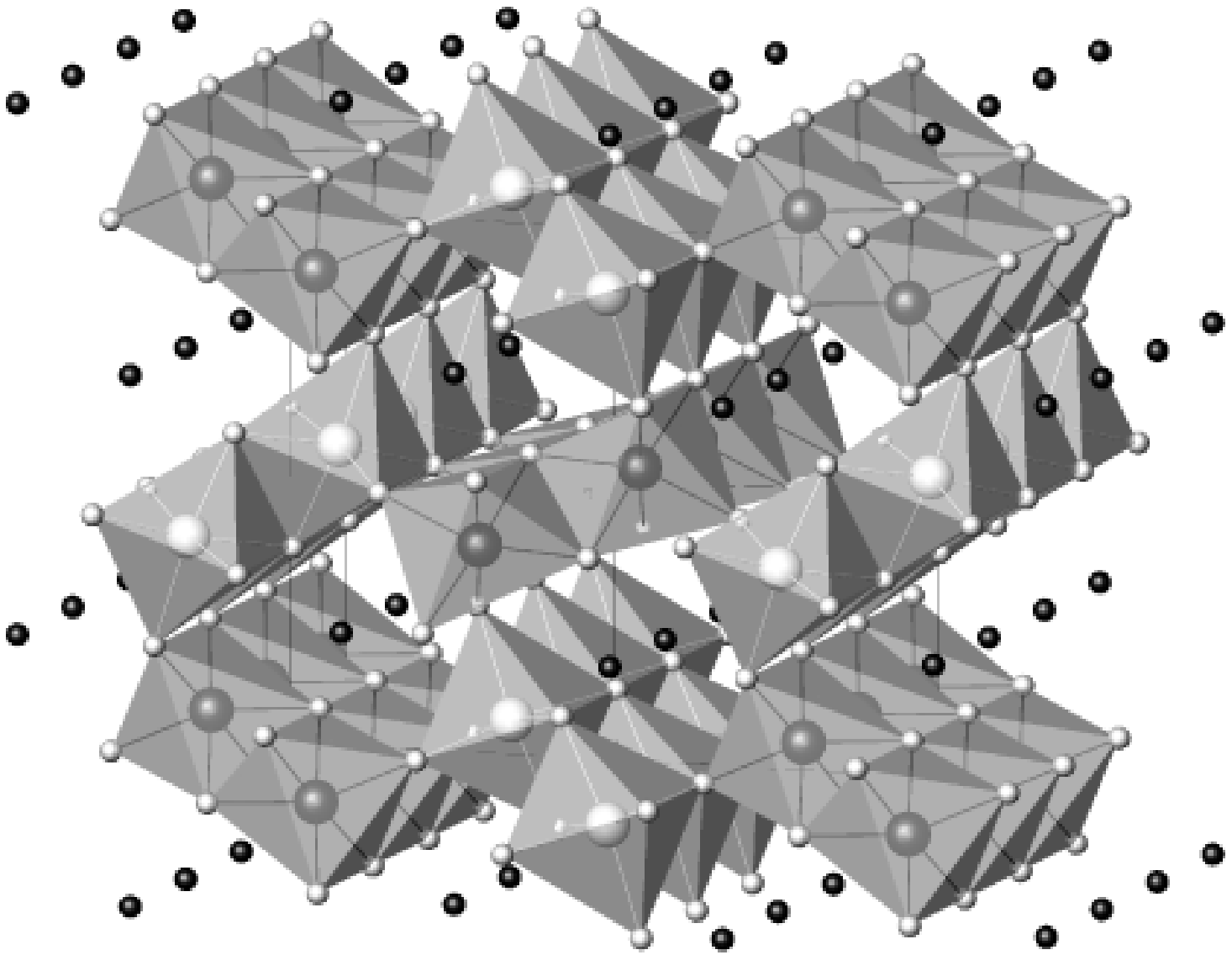}
\caption{End-on (top) and inclined (bottom) views along the $c$-axis of
the CaV$_2$O$_4$ structure showing the V zigzag chains with the V atoms in
distorted edge- and corner-sharing octahedral coordination by oxygen.}
\label{structure1}
\end{figure}

\begin{figure}[t]
\includegraphics[width=3in]{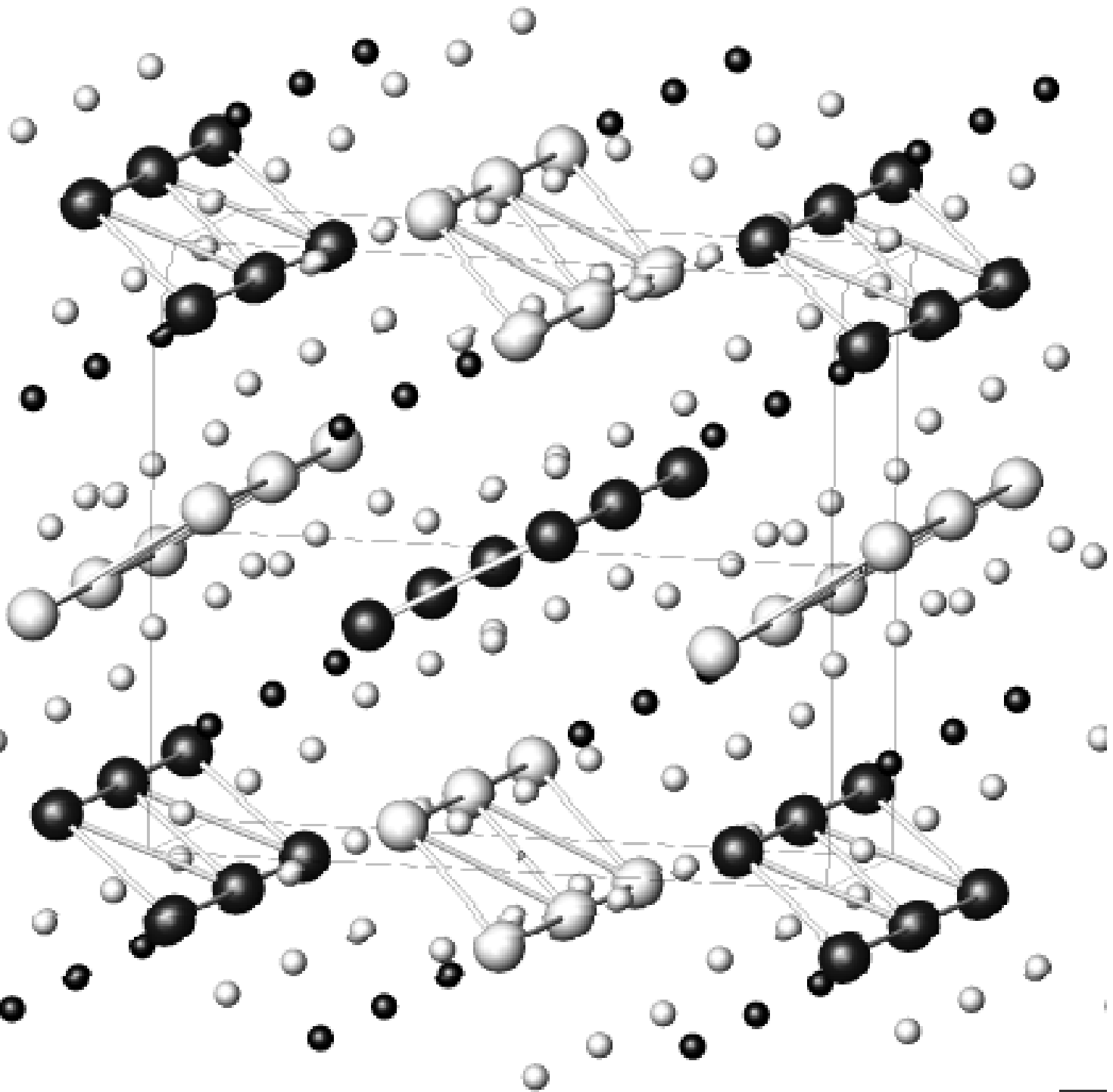}\\
\includegraphics[width=3in]{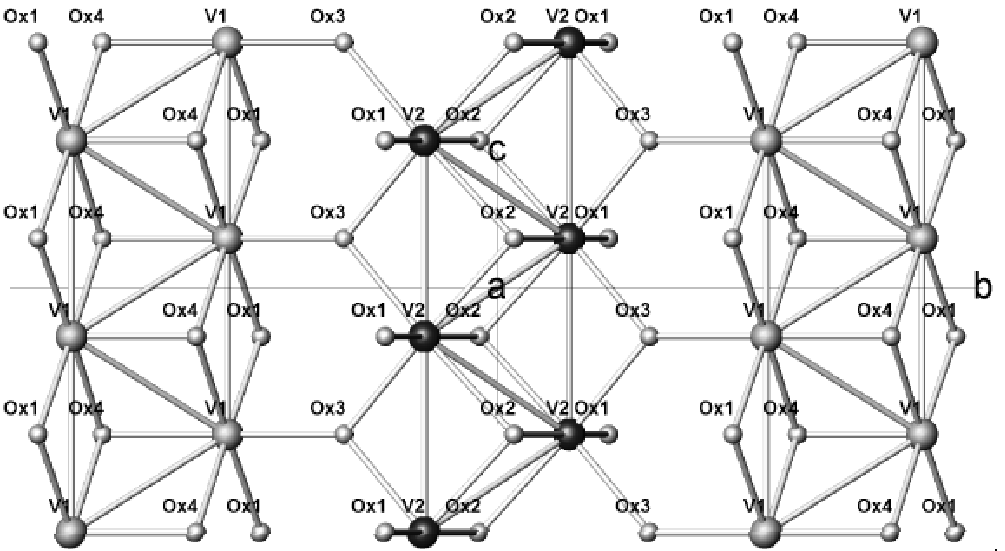}\\
\includegraphics[width=3in]{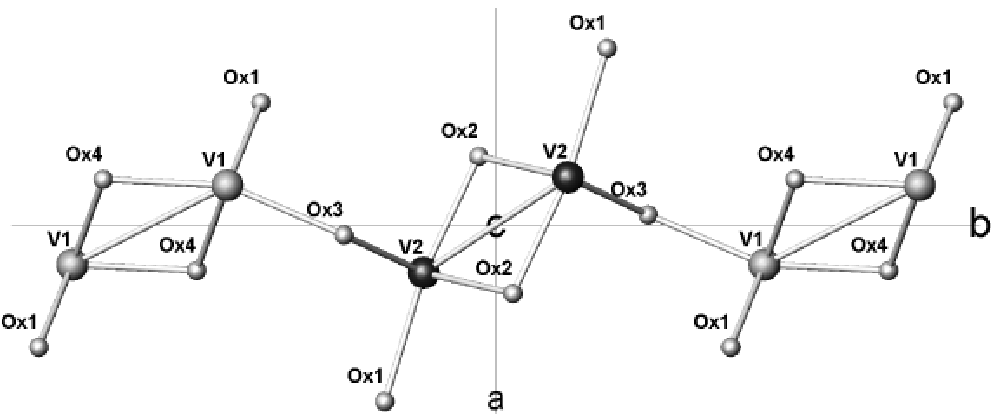}\\
\caption{Top panel: the skeletal structure of CaV$_2$O$_4$ showing the
zigzag V-V chains.  The large spheres represent V atoms, the small dark
spheres Ca atoms, and the small light spheres O atoms.  Middle and bottom
panels: cross-sections of the $b$-$c$ and $a$-$b$ planes, with the V and O
atoms labeled as described in Table~\ref{tabStruct2}.  The Ca atoms have
been omitted for clarity.}
\label{structure2}
\end{figure}

The compound CaV$_2$O$_4$, containing crystallographic V$^{+3}$ spin $S=1$
zigzag chains, has been suggested  as a model experimental system to study the
above chiral gapless phase. \cite{fukushima, kikuchi}  CaV$_2$O$_4$
crystallizes in the CaFe$_2$O$_4$ structure at room temperature\cite{decker,
hastings} with the orthorhombic space group \emph{Pnam} and with all the atoms
in distinct Wyckoff positions 4($c$) ($x,y,1/4$) in the unit cell.  As shown in
Figs.~\ref{structure1} and
\ref{structure2}, two zigzag chains of distorted edge- and corner-sharing VO$_6$
octrahedra occur within the unit cell and run parallel to the 
$c$-axis, with the Ca ions situated in tunnels between the chains.  Two sets of
crystallographically inequivalent V atoms occupy the two zigzag chains,
respectively.  The VO$_6$ octahedra within a zigzag chain share corners with
the octahedra in the adjacent zigzag chain.  Within each zigzag chain, in order to be consistent with our theoretical modeling later in Sec.~\ref{calc} of the paper, the nearest neighbors are \emph{defined} to be those on different legs of the zigzag chain where the NN V-V distance is 3.07~\AA.  The NNN V-V distance (3.01~\AA) is \emph{defined} to be along a leg of the zigzag chain.  The similarity between these two distances in CaV$_2$O$_4$ suggests that $J_1\approx J_2$, which would result in geometrically frustrated AF interactions in this insulating low-dimensional system.\cite{fukushima, kikuchi}

Previous studies on polycrystalline samples of CaV$_2$O$_4$ have offered
contrasting views on the nature of the magnetic ground state.  Magnetic
neutron diffraction measurements on CaV$_2$O$_4$
(Ref.~\onlinecite{hastings}) gave clear evidence for the presence of long-range
antiferromagnetic ordering at 4.2~K (the temperature dependence was not
studied, and the N\'eel temperature was not determined).  A doubled magnetic
unit cell along the $b$ and $c$ directions was found with AF 
propagation vector (0~$\frac{1}{2}$~$\frac{1}{2}$) and three collinear AF
models with the V ordered moments parallel to the $b$-axis were considered. 
Interestingly, the ordered moment was found to be 1.06(6)~$\mu_{\rm B}$/(V
atom), where $\mu_{\rm B}$ is the Bohr magneton.  This value is a factor of two
smaller than the value $gS\mu_{\rm B} = 2.0~\mu_{\rm B}$/(V atom) expected for
a spin $S = 1$ with spectroscopic splitting factor ($g$-factor) $g = 2$. 
Magnetic susceptibility measurements\cite{fukushima, kikuchi} showed a broad maximum at $\sim 250$~K, indicating the onset of strong short-range AF ordering in a low-dimensional spin system upon cooling.  The data also showed a finite value
at the lowest temperatures, indicating that an energy gap for spin excitations
did not occur, consistent with the neutron diffraction measurements.  However,
these magnetic susceptibility data also showed a bifurcation below $\sim 20$~K
between low field (100 Oe) zero-field-cooled and field-cooled  measurements
that was suggestive of a spin-glass-like freezing rather than long-range AF
ordering.  $^{51}$V nuclear magnetic resonance (NMR)  measurements\cite{fukushima, kikuchi} showed a
nuclear spin-lattice relaxation rate $1/T_1 \propto T$ at low temperatures from
2~K to 30~K, of unknown origin, but again indicating  lack of an energy gap for
magnetic excitations.  The authors\cite{fukushima, kikuchi} proposed a {\em
chiral gapless ordered} phase at low temperatures in accordance with theoretical
predictions for a $S=1$ frustrated $XY$ or $XXZ$ chain.  The chiral phase
implies a helical spin arrangement in contrast to the collinear spin models
proposed\cite{hastings} in the neutron diffraction study.  Furthermore, the
observation of a $^{51}$V nuclear resonance at the normal
$^{51}$V Larmor frequency\cite{fukushima, kikuchi} at temperatures at and below
4~K is not consistent with the long-range antiferromagnetic ordering found at
4~K from the neutron diffraction measurements,\cite{hastings} since such
ordering produces a very large static local magnetic field of order 20~T at the
positions of the V nuclei.

In order to resolve the above inconsistencies regarding the magnetic
ground state of CaV$_2$O$_4$ and to search for interesting physics in this
system associated with possible geometric frustration within the zigzag spin
chains, we have for the first time (to our knowledge) grown single crystals of
this compound, and report herein crystal structure,  static magnetization and magnetic
susceptibility $\chi(T)$, ac magnetic susceptibility $\chi_{\rm ac}(T)$, heat
capacity $C_{\rm p}(T)$, and anisotropic linear thermal expansion $\alpha_i(T)$
($i = x,y,z$) measurements over the temperature $T$ range 1.8 to 350~K on
polycrystalline and single crystal samples.   Our $\chi(T)$ and $\chi_{\rm
ac}(T)$ measurements do not show any signature of a spin-glass transition
around 20~K as previously reported.\cite{fukushima, kikuchi}  We instead
observe long-range antiferromagnetic (AF) ordering at sample-dependent N\'eel
temperatures $T_{\rm N}\approx 51$--76~K\@.  

We have recently reported elsewhere the results of $^{17}$O NMR measurements
on a polycrystalline sample of CaV$_2$O$_4$ and find a clear signature of AF
ordering at $T_{\rm N}\approx 78$~K\@.\cite{zong}  We find no evidence of a
$^{51}$V NMR signal at the normal Larmor frequency at temperatures
between 4~K and 300~K, in conflict with the above  previous $^{51}$V NMR
studies which did find such a resonance.\cite{fukushima, kikuchi}  In single
crystals, at temperatures below 45~K we do find a \emph{zero-field} $^{51}$V NMR
signal where the $^{51}$V nuclei resonate in the static component of the local
magnetic field generated by the long-range AF order below $T_{\rm N} \approx
70$~K\@.\cite{zong}  The ordered moment at 4.2~K in the crystals was estimated
from the zero-field $^{51}$V NMR measurements to be 1.3(3)~$\mu_{\rm B}$/(V
atom), somewhat larger than but still consistent with the value 1.06(6)~$\mu_{\rm B}$/(V atom) from the above
neutron diffraction measurements.~\cite{hastings}   An energy gap $\Delta$ for
antiferromagnetic spin wave excitations was found with a value
$\Delta/k_{\rm B} =  80(20)$~K in the temperature range 4--45~K, where $k_{\rm
B}$ is Boltzmann's constant.  This energy gap was proposed to arise from
single-ion anisotropy associated with the $S = 1$ V$^{+3}$ ion.  A model for the
antiferromagnetic structure at 4~K was formulated in which the magnetic
structure consists of two substructures, each of which exhibits collinear
antiferromagnetic order, but where the easy axes of the two substructures are
at an angle of 19(1)$^\circ$ with respect to each other.  The \emph{average}
easy axis direction is along the $b$-axis, consistent with our magnetic
susceptibility measurements to be presented here, and with the easy-axis
direction proposed in the earlier neutron diffraction
measurements.\cite{hastings}  Our magnetic neutron diffraction studies of
the antiferromagnetic structure of CaV$_2$O$_4$ single crystals are qualitatively consistent with the NMR analyses; these results together with
high-temperature ($T \leq 1000$~K) magnetic susceptibility measurements
and their analysis are presented elsewhere.\cite{Pieper2008B}

We also find that CaV$_2$O$_4$ exhibits a weak orthorhombic to monoclinic
structural distortion upon cooling below a sample-dependent temperature $T_{\rm
S} = 108$--147~K, discovered from our neutron and x-ray diffraction measurements to be
reported in detail elsewhere.\cite{bella}  In our two \emph{annealed single crystals} only, anomalies in the heat capacity and thermal expansion are also found at $T_{\rm S1} \approx 200$~K\@.  From high-energy x-ray diffraction measurements reported here, we find that in one of the crystals the anomaly is most likely primarily due to the metal-insulator and structural transitions in V$_2$O$_3$ impurity phase that grows coherently in the crystal when it is annealed.  In the other crystal, we still find a small anomaly in the heat capacity at $T_{\rm S1}$ but where the transition in the V$_2$O$_3$ impurity phase is at much lower temperature.  Hence we infer that there is an intrinsic transition in our two annealed CaV$_2$O$_4$ crystals at $T_{\rm S1}$ with an unknown origin.  We speculate that this transition may be the long-sought chiral ordering transition envisioned by Villain\cite{Villain1977} that was mentioned above.

From our inelastic neutron scattering results to be published
elsewhere,\cite{Pieper2008} we know that the magnetic character of CaV$_2$O$_4$
is quasi-one-dimensional as might be inferred from the crystal structure.  The
largest dispersion of the magnetic excitations is along the zigzag chains,
which is along the orthorhombic $c$-axis direction, with the dispersion along
the two perpendicular directions roughly a factor of four smaller.   This indicates that the exchange interactions perpendicular to the zigzag chains are roughly an order of magnitude smaller than within a chain.  We
therefore analyze the magnetic susceptibility results here in terms of theory
for the $S = 1$ $J_1$-$J_2$ linear Heisenberg chain, where $J_1$($J_2$) is the
\mbox{(next-)nearest}-neighbor interaction along the chain.  With respect to the
interactions, this chain is topologically the same as a zigzag chain where
$J_1$ is the nearest-neighbor interaction between spins in the two different
legs of the zigzag chain and $J_2$ is the nearest-neighbor interaction between
spins within the same leg.  We utilize exact diagonalization to calculate the
magnetic susceptibility and magnetic heat capacity for chains containing up to
14 spins $S = 1$, and quantum Monte Carlo simulations of the magnetic
susceptibility and heat capacity for chains of 30 and 60 spins.  We  obtain estimates
of $J_1$, $J_2/J_1$, the temperature-independent orbital susceptibility
$\chi_0$, and the zero-temperature spin susceptibilities in CaV$_2$O$_4$ from
comparison of the theory with the experimental $\chi(T)$ data near room
temperature.  Remarkably, we find that one of the two exchange constants is very small compared to the other near room temperature, as opposed to $J_1/J_2 \approx 1$ that is expected from the crystal structure.  Thus, with respect to the magnetic interactions, the zigzag crystallographic chain compound acts instead like a linear $S=1$ Haldane spin chain compound.  In Ref.~\onlinecite{Pieper2008B}, we propose that partial orbital ordering is responsible for this unexpected result, and suggest a particular orbital ordering configuration.  In Ref.~\onlinecite{Pieper2008B}, we also deduce that below $T_{\rm S}\sim 150$~K, the monoclinic distortion results in a change in the orbital ordering that in turn changes the nature of the spin interactions from those of a Haldane chain to those of a $S =1$ two-leg spin ladder.  Here we also compare the theory for the magnetic heat capacity with the results of our heat capacity experiments.  We estimate the coupling $J_\perp$ between these chains that leads to the long-range AF order at $T_{\rm N}$.  

The remainder of this paper is organized as follows.  The synthesis and
structural studies are presented in Sec.~\ref{struct}.  The 
magnetization, magnetic susceptibility, heat capacity and thermal expansion
measurements are presented in Sec.~\ref{measure}.  In Sec.~\ref{calc} we
consider the origin of the heat capacity and thermal expansion anomalies at $T_{\rm S1} \approx 200$~K\@.  We then analyze the $\chi(T)$ data in terms of the predictions of exact diagonalization calculations and quantum Monte Carlo simulations to obtain $J_1$, $J_2/J_1$ and $\chi_0$.   Using the same $J_1$ and  $J_2/J_1$ parameters, we compare the predicted behavior of the magnetic heat capacity with the experimentally observed heat capacity data.  We also obtain an estimate of the interchain coupling $J_\perp$ giving rise to long-range AF order at $T_{\rm N}$.  A summary of our results is given in Sec.~\ref{summary}.

\section{\label{struct}Synthesis, single crystal growth, and crystal structure of ${\rm{\mathbf{CaV_2O_4}}}$}

\subsection{Synthesis and Crystal Growth}

Polycrystalline CaV$_2$O$_4$ was synthesized via solid state reaction by
first mixing V$_2$O$_3$ (99.995\%, MV Labs) with CaCO$_3$  (99.995\%,
Aithaca Chemicals) or CaO obtained by calcining the CaCO$_3$ at
1100~$^{\circ}$C\@.  The chemicals were ground inside a He glove-box,
pressed and sintered at 1200~$^{\circ}$C for 96 hours in flowing
4.5\%H$_2$-He, as well as in sealed quartz tubes when using CaO, with
intermediate grindings.  Phase purity was confirmed by powder x-ray
diffraction (XRD) patterns obtained using a Rigaku Geigerflex
diffractometer with Cu K$\alpha$ radiation in the angular range $2\theta =
10$--90$^{\circ}$ accumulated for 5~s per $0.02^{\circ}$ step.  
Thermogravimetric analysis (TGA) at 800~$^{\circ}$C using a Perkin Elmer
TGA 7 was used to check the oxygen content by oxidizing the sample to
CaV$_2$O$_6$.  A typical oxygen content of CaV$_2$O$_{3.98\pm0.05}$ was
determined, consistent with the initial stoichiometric composition
CaV$_2$O$_4$. 

\begin{figure}[t]
\includegraphics[width=2.8in]{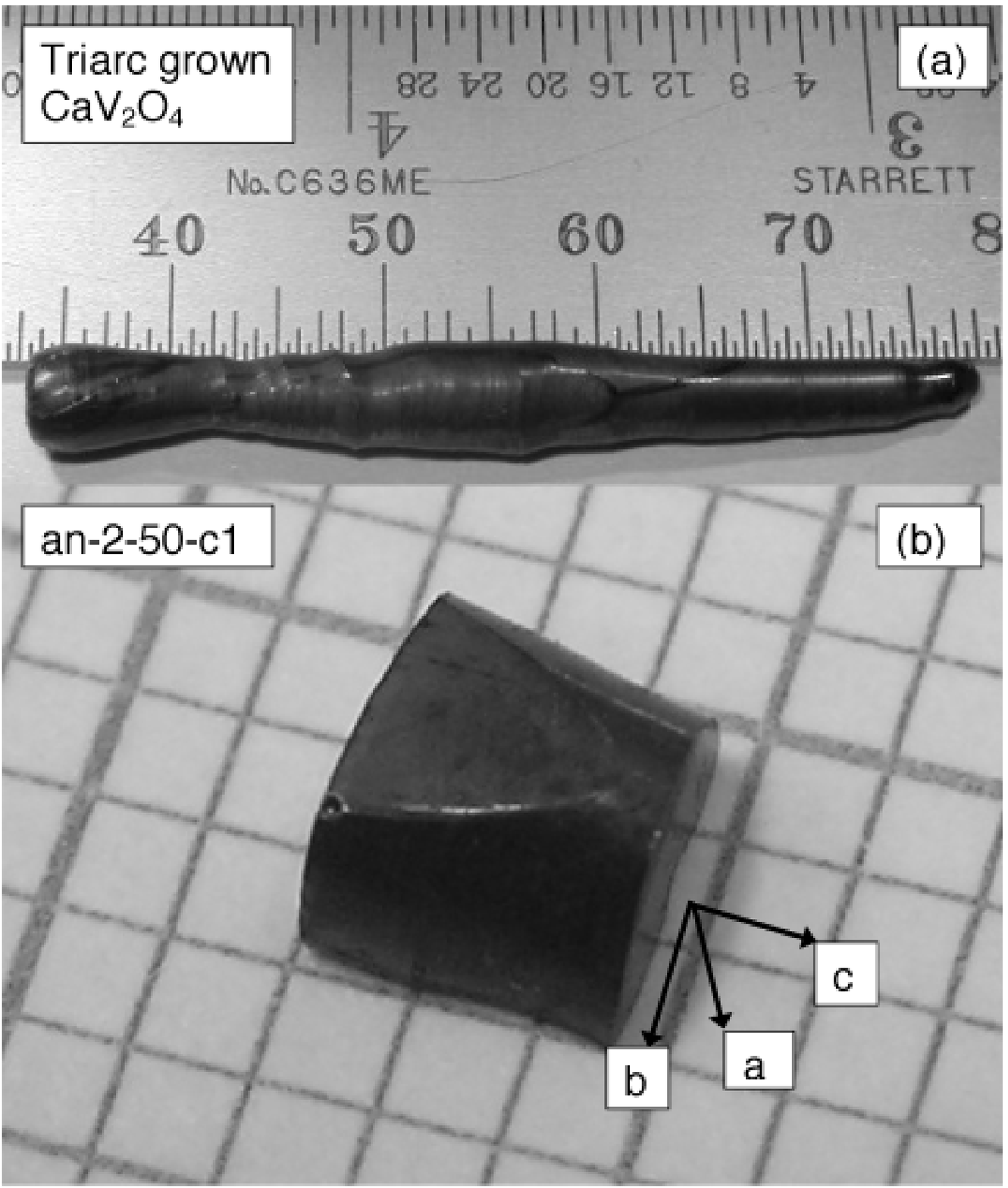}\\
\includegraphics[width=2.8in]{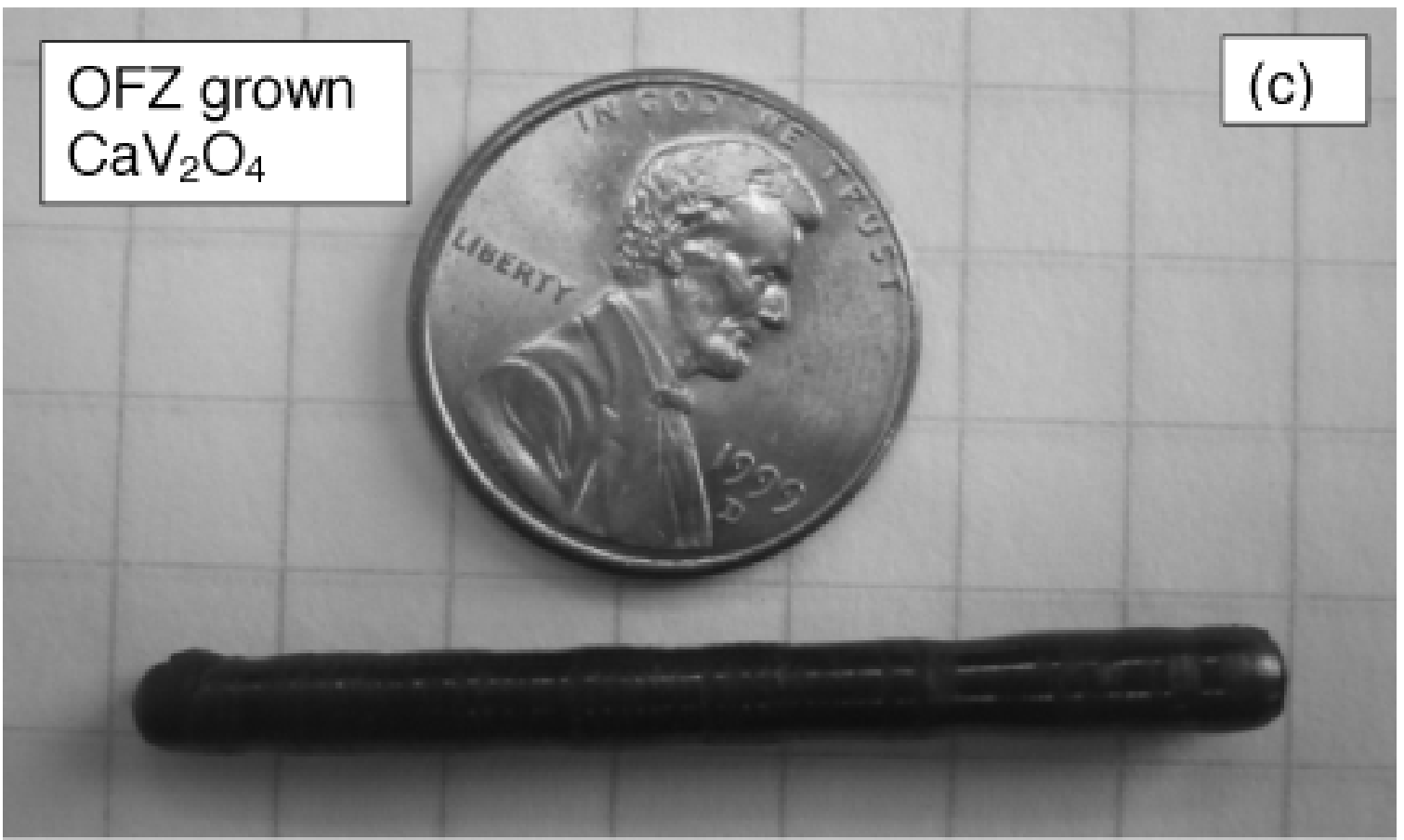}
\caption{Single crystals of CaV$_2$O$_4$ grown using (a), (b) 
 a triarc furnace (a numbered division on the scale is 1 mm) and (c) an
optical floating zone furnace (compared with a U.S. penny).}
\label{crystal-pic}
\end{figure}

CaV$_2$O$_4$ was found to melt congruently in an Ar arc furnace with
negligible mass loss by evaporation.  Therefore crystal growth was
attempted by pulling a crystal from the melt in a triarc
furnace (99.995\% Ar) using a tungsten seed rod.\cite{mpc}  The triarc
furnace was custom made for us by Materials Research Furnaces, Inc.  Using
15--20~g premelted buttons of CaV$_2$O$_4$, pulling rates of 0.2--0.5~mm/min
were used to grow ingots of about 3--6~mm diameter and 3.0--4.7~cm length.  The
length of the ingot was limited by contraction of the molten region as power
was lowered to control the crystal diameter.  Obtaining a single grain was
difficult because small fluctuations in the arcs coupled with high mobility of
the CaV$_2$O$_4$ melt easily caused nucleation of new grains.  Out of multiple
growth runs, a reasonably large single grain section could be cut out of one of
the ingots.  The as-grown ingot from the triarc furnace and a single
crystal isolated and aligned from it are shown in
Figs.~\ref{crystal-pic}(a) and (b), respectively.  Due to the tendency for
multiple   nucleations in the triarc furnace, an optical floating zone (OFZ)
furnace was subsequently used for crystal growth.\cite{mpc}  Growth rates and Ar
atmosphere flow rates were optimized to successfully grow large crystals
of 4--5~mm diameter and 4--5~cm length starting from sintered
polycrystalline rods with masses of 8--10~g.  An as-grown rod from the
OFZ furnace is shown in Fig.~\ref{crystal-pic}(c).

Powder XRD of crushed sections from the triarc grown ingots
as well as from the OFZ grown crystals showed single phase CaV$_2$O$_4$.  Laue
x-ray diffraction patterns of a single-grain section confirmed its
single-crystalline character and the crystal was found to grow
approximately along its crystallographic $c$-axis.  The crystals were
oriented and cut to obtain faces aligned perpendicular to the principal
axes.  They were measured as grown (only for the triarc grown crystals) as
well as after annealing in flowing 5\%H$_2$-He gas at 1200~$^{\circ}$C
for up to 96 hours.

\subsection{Powder and Single Crystal X-ray and Neutron Diffraction Measurements}

\begin{figure*}[t]
\includegraphics[width=4in]{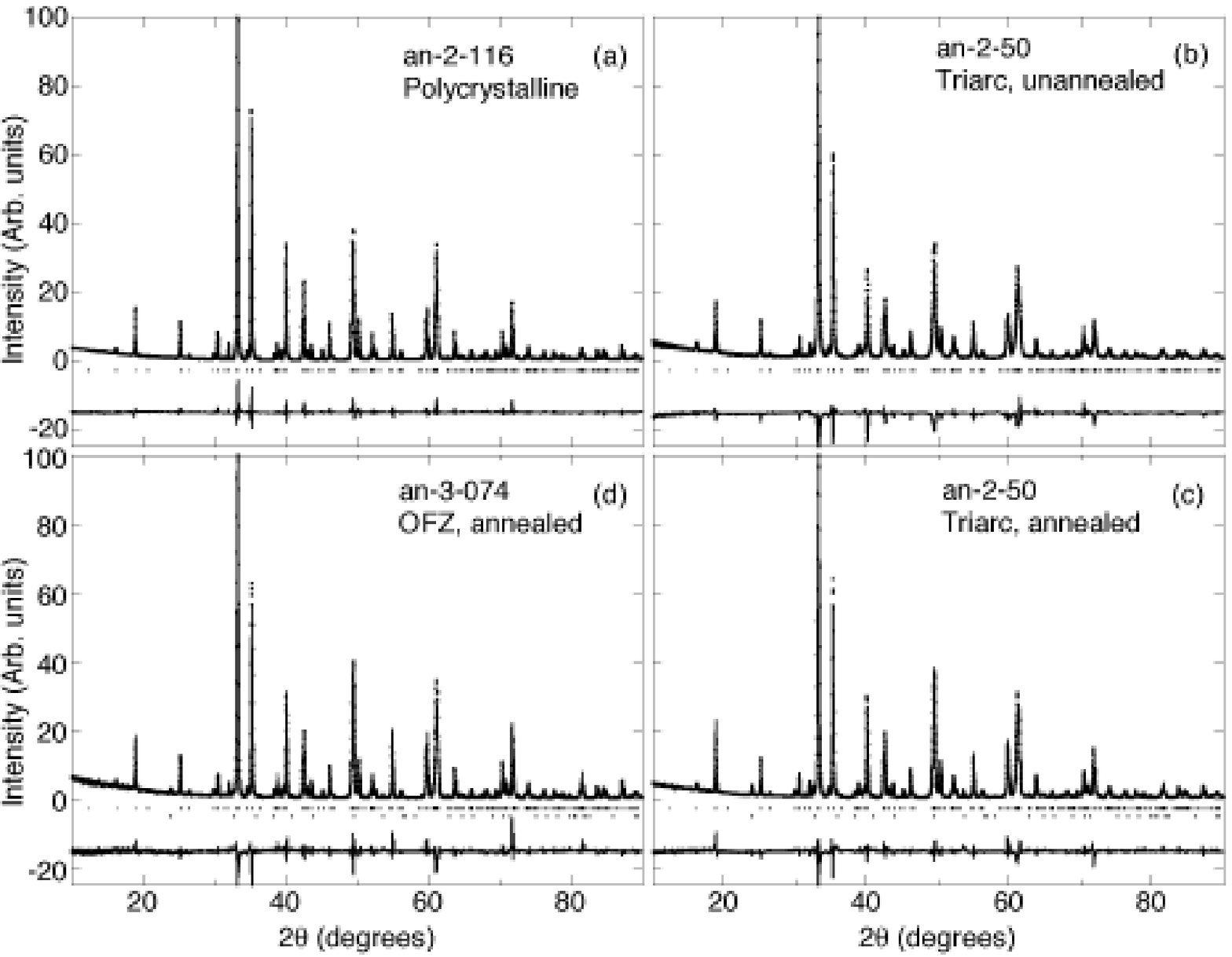}
\caption{Rietveld refinement of room temperature powder XRD data of
CaV$_2$O$_4$ showing I$_{\rm obs}$ (+) I$_{\rm calc}$ (.), difference
(--), and peak positions (\textbar) for (a) a polycrystalline sample, (b)
an as-grown triarc crystal, (c) an annealed triarc crystal, and (d) an annealed
optical floating zone (OFZ) crystal.  The annealed single crystal samples
contain small XRD peaks from $\sim 1$--2 mol\% of V$_2$O$_3$, shown as the
lower sets of peak position markers.} 
\label{xrd}
\end{figure*}

Rietveld refinements of the powder x-ray diffraction data obtained at room
temperature were carried out using the program DBWS9807a.\cite{dbws} The
refined powder XRD patterns from a polycrystalline sample and crushed
pieces of the triarc and optical floating zone grown crystals are shown in
Fig.~\ref{xrd} and the refinement results are presented in
Tables~\ref{tabStruct}, \ref{tabStruct2}, and~\ref{tabStruct3}.  The XRD of the
powdered \emph{annealed} single crystal samples showed a trace amount ($\sim$
1--2 mol\%) of V$_2$O$_3$.  It is curious that no trace was found of V$_2$O$_3$ impurity phase in the as-grown crystals, and that this impurity phase only formed after annealing the crystals.  From the refinement results, the structural parameters remain relatively unchanged between polycrystalline samples prepared by the solid state route and both as-grown and annealed single crystals grown from the melt. 

\begin{figure}[t]
\includegraphics[width=3in]{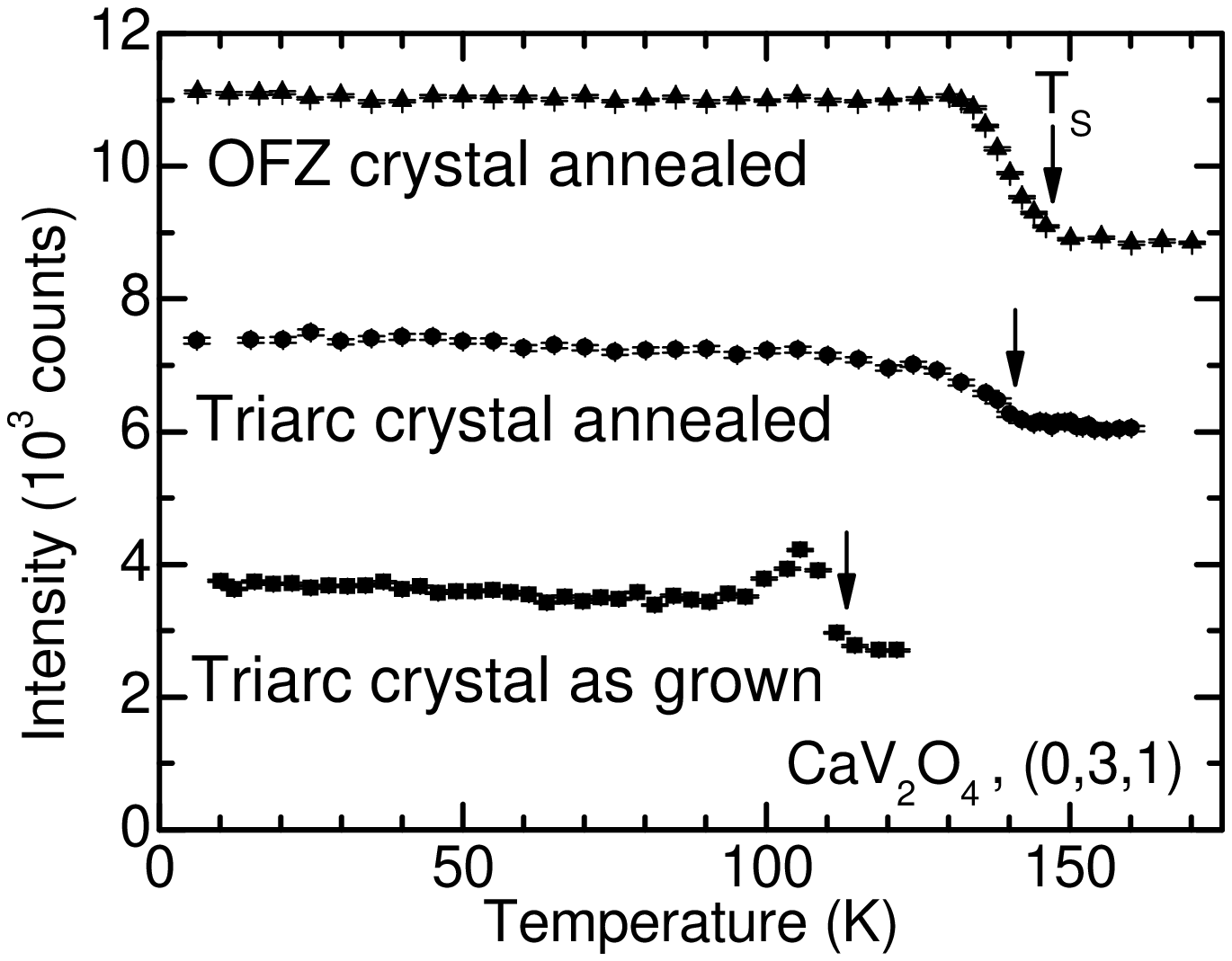}
\caption{Temperature variation of intensity of the (0 3 1) structural
Bragg peak across the orthorhombic to monoclinic structural transition
temperature ($T_{\rm S}$) in single crystal samples of CaV$_2$O$_4$
measured by neutron diffraction.  The (0~3~1) peak is present in both the
orthorhombic and monoclinic phases.} 
\label{neutron031}
\end{figure}

Single-crystal neutron diffraction data were collected on the four-circle
diffractometer E5 at the BERII reactor of the Helmholtz-Zentrum Berlin, Germany.  A pyrolytic graphite monochromator was used to select
the neutron wavelength $\lambda = 2.36$~\AA.  Second order contamination
was suppressed below $10^{-3}$ of first order by a pyrolytic graphite filter. 
Bragg reflections of CaV$_2$O$_4$ were measured with a two-dimensional position
sensitive $^3$He detector, $90\times 90$~mm$^2$ in area.  The sample was
mounted in a closed-cycle refrigerator, where the temperature was
controlled between 290~K and 6~K\@.  A structural phase transition at
temperature $T_{\rm S}$ from the high temperature orthorhombic structure to a
low temperature monoclinic structure was found.\cite{bella}  This transition is
reflected in Fig.~\ref{neutron031} by a sudden change in the (0~3~1) Bragg peak
intensity which occurs at a temperature $T_{\rm S} \approx 112$~K for the
as-grown triarc crystal, and $\approx 141$~K and $\approx 147$~K for the
annealed triarc and OFZ-grown crystals, respectively.  Due to twinning the orthorhombic (0 3 1) reflection splits below the structural phase transition into the (0 3 1) and (0 $\bar{3}$ 1) monoclinic reflections. The total integrated intensity at this position increases at $T_{\rm S}$ because of the increased mosaic which results in a reduction of the extinction effect.  The peak in the intensity at 105~K for the as-grown triarc crystal is
an experimental artifact due to multiple scattering.  The lattice parameters of
the low temperature monoclinic phase differ very little from the orthorhombic
phase and the monoclinic angle $\alpha \approx 89.3^{\circ}$ is close to
$90^{\circ}$.  This result and the smoothly varying signatures in the
thermodynamic properties suggest that the structural transition is of second
order and involves a small distortion of the orthorhombic structure.  Full details of the neutron and x-ray diffraction structural measurements and results will be presented elsewhere.\cite{bella}

A higher temperature anomaly in the temperature dependence of the lattice parameters of a powder sample was observed by x-ray diffraction over a temperature range of
175--200~K\@.  This transition with $T_{\rm S1} \approx 200$~K was also observed
in the magnetic susceptibility, thermal expansion, and heat capacity
measurements of two \emph{annealed single crystals} as described in Sec.~\ref{measure} below.  In the next section we investigate whether there is a structural aspect to this phase transition.

\begin{table*}
\caption{\label{tabStruct}  Structure parameters at room temperature for CaV$_2$O$_4$ forming
in the CaFe$_2$O$_4$ structure, refined from powder XRD data. Space
Group: $Pnam$ (\#62); $Z = 4$; Atomic positions: 4(c), ($x, y, 1/4$);
Profile: Pseudo-Voigt. The overall isotropic thermal parameter $B$ is
defined within the temperature factor of the intensity as $e^{-2B \sin^2
\theta/ \lambda^2}$.}
\begin{ruledtabular}
\begin{tabular}{llcccccc}
  Sample & Synthesis  & $a$ (\AA)& $b$ (\AA) & $c$
(\AA)&  $B$ (\AA$^2$) & $R_{\rm wp}(\%)$ & $R_{\rm p}$ (\%)\\
\hline
an-2-116 &  1200~$^\circ$C solid state & 9.2064(1) & 10.6741(1) &  
3.0090(1) &  2.33(4) & 11.17 & 8.24\\
an-2-50& Triarc as grown & 9.2241(11) & 10.6976(13) & 3.0046(4) &  
1.72(5) & 12.05 & 9.27\\
an-2-50 & Triarc annealed 1200~$^\circ$C & 9.2054(3) & 10.6748(3) &  
3.0042(1) &1.58(5) & 14.54 & 10.95 \\
an-3-074& OFZ annealed 1200~$^\circ$C & 9.2089(2) & 10.6774(3) &  
3.0067(1) &0.75(5) &17.46 & 12.77\\
\end{tabular}
\end{ruledtabular}
\end{table*}

\begin{table*}
\caption{\label{tabStruct2}  Atomic positions $(x,y,1/4)$ for  
CaV$_2$O$_4$ obtained by Rietveld refinement of powder XRD data at room temperature for four
samples. }
\begin{ruledtabular}
\begin{tabular}{lcccc}
Sample Number  & an-2-116 & an-2-50-c1 & an-2-50-c1 & an-3-074 \\
Synthesis & Solid State & Triarc   & Triarc & OFZ \\
  & (1200 $^\circ$C) &  (as grown)  & (annealed)\footnotemark[1] &   
(annealed)\footnotemark[1] \\
& $x,\ y$ & $x,\ y$ & $x,\ y$ & $x,\ y$\\
\hline
Ca & 0.7550(3), 0.6545(2) & 0.7562(4), 0.6536(3) & 0.7542(4),  
0.6544(3) & 0.7536(4), 0.6550(3)\\
V1 & 0.4329(2), 0.6117(1) & 0.4320(3), 0.6120(2) & 0.4336(3),  
0.6120(2) & 0.4331(3), 0.6114(2)\\
V2 & 0.4202(2), 0.1040(1) & 0.4204(3), 0.1041(2) & 0.4200(3),  
0.1043(2) & 0.4209(3), 0.1043(2)\\
O1 & 0.2083(6), 0.1615(5) & 0.2128(8), 0.1593(7) & 0.2049(8),  
0.1603(8) & 0.2074(9), 0.1635(9)\\
O2 & 0.1176(5), 0.4744(5) & 0.1157(7), 0.4745(8) & 0.1144(7),  
0.4756(8) &  0.1181(9), 0.4738(9)\\
O3 & 0.5190(7), 0.7823(5) & 0.5153(11), 0.7812(7) & 0.5166(0),  
0.7806(7) & 0.5169(11), 0.7797(8) \\
O4 & 0.4203(6), 0.4270(5) & 0.4207(8), 0.4282(7) & 0.4244(8),  
0.4325(7) & 0.4280(9), 0.4251(9)\\
\footnotetext[1]{Two-phase sample containing $\sim 1$\% V$_2$O$_3$ as  
determined from Rietveld refinement.}
\end{tabular}
\end{ruledtabular}
\end{table*}

\begin{table*}
\caption{\label{tabStruct3}  Bond angles V--O--V and bond lengths V--V for
CaV$_2$O$_4$ at room temperature refined from powder XRD data and calculated using Atoms for
Windows, version~5.0.  The V$_1$--O and V$_2$--O bond lengths varied from
1.92~\AA\ to 2.08~\AA.  The accuracy of the bond angles calculated is
$\pm 0.1$~$^{\circ}$.} 
\begin{ruledtabular}
\begin{tabular}{lcccc}  Sample Number  & an-2-116 & an-2-50-c1 &
an-2-50-c1 & an-3-074 \\  Synthesis & Sintered powder & Triarc   & Triarc
& Optical float zone \\
 & (1200 $^\circ$C) &  (as grown)  & (annealed) &  (annealed) \\
\hline V$_1$--O$_1$--V$_1$ (NN)($^{\circ}$) & 93.9 & 92.9 & 95.0 & 93.7\\ 
V$_1$--O$_4$--V$_1$ (NN)($^{\circ}$) & 93.6 & 93.0 & 94.4 & 96.6\\ 
V$_1$--V$_1$ (NN)(\AA) & 3.009 & 3.005 & 3.004 & 3.004\\
V$_1$--O$_4$--V$_1$ (NNN)($^{\circ}$) & 99.3 & 99.9 & 101.8 & 100.3\\
V$_1$--V$_1$ (NNN)(\AA) & 3.078 & 3.094 & 3.077 &  3.071\\
V$_2$--O$_2$--V$_2$ (NN)($^{\circ}$) & 93.1 & 93.0 & 93.7 & 92.5 \\
V$_2$--O$_3$--V$_2$ (NN)($^{\circ}$) & 96.8 & 95.6 & 95.6 & 97.5\\ 
V$_2$--V$_2$ (NN)(\AA)  & 3.009 & 3.005 & 3.004 & 3.004 \\
V$_2$--O$_3$--V$_2$ (NNN)($^{\circ}$) & 97.3 & 98.0 & 98.3 & 97.1 \\
V$_2$--V$_2$ (NNN)(\AA) & 3.058 & 3.062 & 3.062 & 3.055 \\
V$_1$--O$_1$--V$_2$ ($^{\circ}$)\footnotemark[1] & 121.7 & 122.9 & 121.8 
& 121.0 \\ V$_1$--V$_2$ (\AA)\footnotemark[1] & 3.583  & 3.581  & 3.582 
& 3.589 \\ V$_1$--O$_3$--V$_2$ ($^{\circ}$)\footnotemark[2] & 131.6 &
132.2 & 132.2 &  132.2\\ V$_1$--V$_2$ (\AA)\footnotemark[2] & 3.647 &
3.652 & 3.652 & 3.643\\
\footnotetext[1]{Interchain angles and distances parallel to $a$-axis}
\footnotetext[2]{Interchain angles and distances parallel to $b$-axis}
\end{tabular}
\end{ruledtabular}
\end{table*}

\subsection{\label{HEXRD}High Energy X-ray Diffraction Measurements on Annealed C\lowercase{a}V$_{\bm{2}}$O$_{\bm{4}}$ Single Crystals}

In order to unambiguously determine the crystallographic 
structure of CaV$_{2}$O$_{4}$ at various temperatures, to 
characterize structural phase transitions, and to check the 
crystal perfection, high-energy x-ray diffraction measurements 
($E$~= 99.43~keV) using an area detector were performed on two 
annealed single crystals at the Advanced Photon Source at Argonne National Laboratory.  At this high energy, x-rays probe the bulk of 
a crystal rather than just the near-surface region and, by 
rocking the crystal about both the horizontal and vertical axes perpendicular to the incident x-ray beam, 
an extended range of a chosen reciprocal plane can be 
recorded.\cite{Kreyssig07} For these measurements, a crystal was 
mounted on the cold-finger of a closed-cycle refrigerator 
surrounded by the heat shield and vacuum containment using Kapton 
windows to avoid extraneous reflections associated with Be or the 
aluminum housing.  Two orientations of the crystal, with either 
the [001] or [100] direction parallel to the incident beam, were 
studied allowing the recording of the ($hk0$) or ($0kl$) 
reciprocal planes.  For each data set, the horizontal angle, 
$\mu$, was scanned over a range of $\pm$2.4~deg for each value of 
the vertical angle, $\eta$, between $\pm$2.4~deg with a step size 
of 0.2~deg. The total exposure time for each frame was 338~sec. 
The x-ray diffraction patterns were recorded with different intensities of the 
incident beam that were selected by attenuation to increase the dynamic 
range to a total of $10^7$ counts. A beam size of $0.3 \times 
0.3$~mm$^2$ was 
chosen to optimize the intensity/resolution condition and to 
allow probing different sections of the crystal by stepwise 
translations of the crystal in directions perpendicular to the incident beam.

\begin{figure*}[t]
\includegraphics[width=4in]{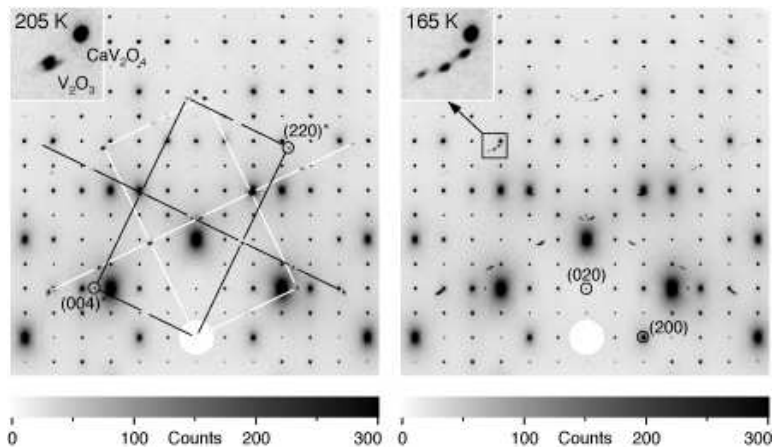}
\caption{\label{fig:xfig1} High-energy x-ray diffraction patterns 
of the annealed triarc CaV$_2$O$_4$ crystal (an-2-50-c1), oriented 
with the ($hk0$) plane coincident with the scattering plane at 
$T$~= 205~K (left panel) and 165~K (right panel). The white 
circles in the lower center of each pattern depict the excluded 
areas around the primary x-ray beam direction.  Several peaks 
corresponding to the main phase, CaV$_2$O$_4$, as well as the 
coherently oriented second phase, V$_2$O$_3$, are labeled by 
indices ($hkl$) and ($hkl$)$^{*}$, respectively. For V$_2$O$_3$, the hexagonal Miller indices for the rhombohedral lattice are used.  The insets of 
both panels display enlarged regions of the diffraction pattern 
to highlight the rhombohedral-to-monoclinic transition for 
V$_2$O$_3$.}
\end{figure*}

\begin{figure}
\includegraphics[width=3in]{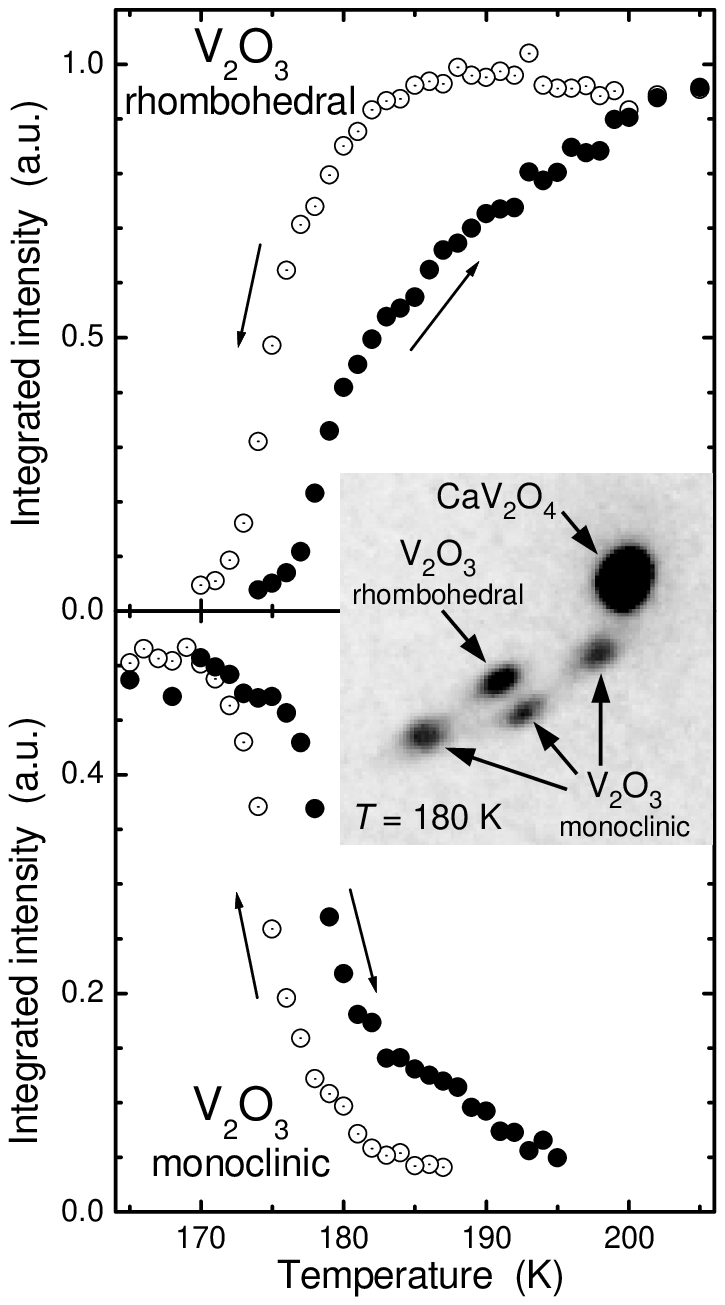}
\caption{\label{fig:xfig2} Integrated intensity of selected 
reflections in the high-energy x-ray diffraction pattern related 
to the rhombohedral (top panel) and monoclinic (bottom panel) phases of V$_2$O$_3$ as 
a function of temperature for the annealed triarc 
crystal (an-2-50-c1).  The direction of the temperature change is indicated by 
arrows.  The inset displays the pattern observed at 180~K (with 
increasing temperature) for the (220) reflection from V$_2$O$_3$ 
in the coexistence temperature range.  This region is the same as 
that displayed in the insets of Fig.~\ref{fig:xfig1}.}
\end{figure}

\subsubsection{\label{AnnTriarc}Annealed Triarc-Grown Crystal}

Figure~\ref{fig:xfig1} shows the ($hk0$) diffraction plane of the annealed 
triarc sample an-2-50-c1 at 205~K\@.  The reciprocal space image 
reveals well-defined diffraction spots that correspond to the 
``primary'' CaV$_2$O$_4$ lattice, as well as spots that can be 
indexed to an impurity phase inclusion of V$_{2}$O$_{3}$ coherently 
oriented with respect to the CaV$_2$O$_4$.  No additional 
reflections were observed.  Indeed, we find two coherent 
twins of V$_{2}$O$_{3}$ related by an inversion across a mirror 
plane of the CaV$_2$O$_4$ lattice as depicted by the black and 
white rectangles in the left panel of Fig.~\ref{fig:xfig1}.  By 
comparing the integrated intensities of reflections from the two 
phases, we estimate that V$_{2}$O$_{3}$ comprises a volume 
fraction of approximately 1--2 percent of the sample.   This is in 
excellent agreement with the result of the x-ray diffraction 
analysis of the polycrystalline sample prepared from the same 
annealed crystal that was described above.  The volume fraction of V$_{2}$O$_{3}$ 
varies only slightly in different parts of the crystal probed by 
scanning the x-ray beam over the crystal.  This indicates that 
the inclusions of V$_{2}$O$_{3}$ are approximately equally 
distributed over the volume of the crystal.

Upon lowering the temperature of the crystal to 165~K, below  
$T_{\rm{S1}} \sim 200$~K, we observe changes in the V$_{2}$O$_{3}$ 
structure consistent with the known first-order rhombohedral-to-monoclinic 
structural transition at 170~K (measured on heating).
\cite{Whan70, Honig1975}  In particular, the upper left 
corners of both panels of Fig.~\ref{fig:xfig1} show enlarged 
views of the region near the ($\overline{3}$80) reflection from 
CaV$_2$O$_4$ and the (220) reflection (in hexagonal notation) for 
the rhombohedral lattice of V$_2$O$_3$. Below $T_{\rm{S1}}$ 
the (220) reflection splits into three reflections in the 
monoclinic phase. The temperature dependence of this transition 
is displayed in Fig.~\ref{fig:xfig2}. Here, we note that there is 
a finite range of coexistence between the rhombohedral and 
monoclinic phases of V$_2$O$_3$ (see the inset to 
Fig.~\ref{fig:xfig2}) and the transition itself has a hysteresis 
of roughly 5--10~K\@. 

Several points regarding Figs.~\ref{fig:xfig1} and~\ref{fig:xfig2} 
are relevant to our interpretation of the 
specific heat and thermal expansion measurements of the annealed 
triarc crystal (an-2-50-c1) to be presented below in Figs.~\ref{FigCaV2O4_DeltaCp_Shift_All} 
and~\ref{dil}, respectively.  First, we note that over the 
temperature range encompassing the features at $T_{\rm{S1}}\sim 
200$~K, there is no apparent change in the diffraction pattern of 
CaV$_2$O$_4$.  These anomalies are instead strongly correlated with the 
rhombohedral-to-monoclinic transition in V$_2$O$_3$.  We further 
note that the temperature for this latter transition is somewhat higher 
than the accepted value of $\approx 170$~K (determined on warming) found in the 
literature.\cite{Whan70,Keer76,Bao98}  This difference is, 
perhaps, due to the fact that the V$_2$O$_3$ and CaV$_2$O$_4$ 
lattices are coupled, as evidenced by the coherent orientation 
relationship between them, so that strains at the phase 
boundaries come into play and can raise the transition 
temperature.\cite{Yonezawa04}  In addition, it is 
reported\cite{Keer76,Bao98} that deviations of the stoichiometry 
from V$_2$O$_3$ can affect the transition temperature significantly. 

\begin{figure}
\includegraphics[width=2in]{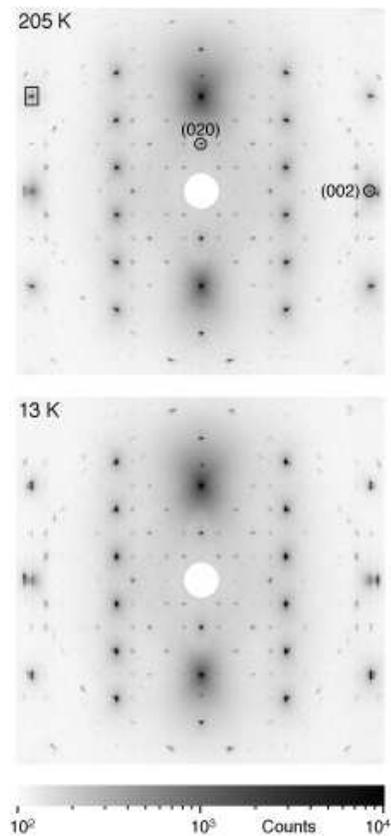}
\caption{\label{fig:xfig4} High-energy x-ray diffraction patterns 
of the ($0kl$)  reciprocal lattice plane of CaV$_2$O$_4$ from the 
annealed triarc crystal (an-2-50-c1) at 205~K (top panel) and 13~K (bottom panel). The white 
circles in the center of the patterns depict the excluded areas 
around the primary x-ray beam direction.  Most of the reflections 
related to CaV$_2$O$_4$ show intensities above 10$^4$ counts (see 
intensity scale). The (020) and (002) reflections of CaV$_2$O$_4$ 
are marked in the top panel. The area bounded by the black rectangle in the top panel depicts the  region close to the orthorhombic (04$\overline{2}$) reflection of CaV$_2$O$_4$, analyzed in Fig.~\ref{fig:xfig5}.}
\end{figure}

\begin{figure}
\includegraphics[width=3in]{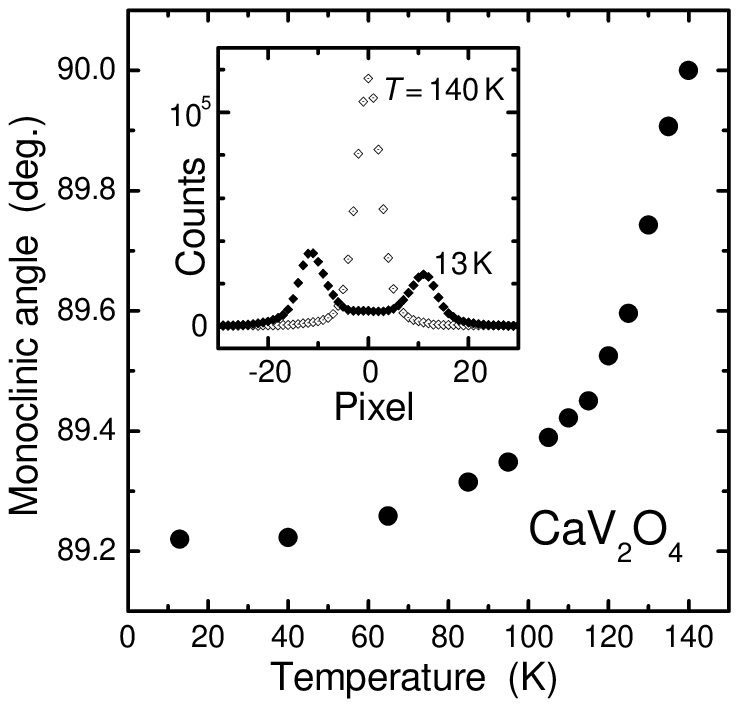}
\caption{\label{fig:xfig5} Temperature dependence of the 
monoclinic angle in the low-temperature phase of CaV$_2$O$_4$, extracted from diffraction data such as shown in the inset for temperatures of 140~K and 
13~K\@. Such diffraction line scans were extracted from high-energy x-ray 
diffraction patterns by summing up the intensity perpendicular to 
the \textbf{b} direction for the sector marked by the rectangle in the top panel of Fig.~\ref{fig:xfig4}.}
\end{figure}

We now turn our attention to changes in the diffraction pattern 
of CaV$_2$O$_4$ associated with anomalies in the heat capacity and thermal expansion measurements at temperatures $T_{\rm{S}} \sim 150$~K in Figs.~\ref{FigCaV2O4_DeltaCp_Shift_All} and~\ref{dil} below, respectively.  The annealed triarc crystal (an-2-50-c1) was reoriented 
so that the CaV$_2$O$_4$ ($0kl$) reciprocal lattice plane was set 
perpendicular to the incident beam.  Figure~\ref{fig:xfig4} shows 
the diffraction patterns obtained at 205~K (above 
$T_{\rm{S}}$) and 13~K (well below $T_{\rm{S}}$).  The 
strong reflections in Fig.~\ref{fig:xfig4} are associated with 
the main CaV$_2$O$_4$ lattice while the weaker diffraction peaks 
are, again, associated with the coherently oriented V$_2$O$_3$ 
second phase.  At low temperatures, we observe a splitting of the 
main reflections that is the signature of the 
orthorhombic-to-monoclinic transition at $T_{\rm{S}}$ for the 
CaV$_2$O$_4$ lattice.  The splitting of reflections associated 
with the transition at $T_{\rm{S1}}$ for V$_2$O$_3$, in this 
reciprocal lattice plane, is not readily observed.

For the low-temperature monoclinic phase of CaV$_2$O$_4$ two possible space 
groups have been considered.\cite{bella}  
The space groups $P~2_1/n~1~1$ and $P~n~1~1$ can be separated by 
testing the occurrence or absence of ($0k0$) reflections with $k$ 
odd, respectively.  The systematic absence of such reflections 
was proven by recording ($hk0$) planes with varying conditions 
to evaluate the sporadic occurrence of these reflections by 
Rengers or multiple scattering.  The space group $P~2_1/n~1~1$ is 
confirmed for the low-temperature phase of the studied 
CaV$_2$O$_4$ crystal.  No changes in the diffraction pattern were 
observed related to the onset of antiferromagnetic order in 
CaV$_2$O$_4$ below $T_{\rm{N}}$ = 69~K\@.

The details of the orthorhombic-to-monoclinic transition at 
$T_{\rm{S}}$ for CaV$_2$O$_4$ are shown in 
Fig.~\ref{fig:xfig5} where we plot the monoclinic distortion angle as a 
function of temperature.  The monoclinic angle was determined 
from the splitting of the peaks along the \textbf{b}-direction 
through the position of the (04$\overline{2}$) reflection. Below 
$T_{\rm{S}} = 138(2)$~K, the monoclinic angle evolves 
continuously, consistent with a second order transition, and 
saturates at approximately 89.2~deg at low temperatures.

\subsubsection{\label{AnnOFZXray}Annealed OFZ-Grown Crystal}

The annealed optical floating zone crystal (an-3-074 OFZ) shows a 
diffraction pattern similar to that of the annealed triarc-grown  
crystal (an-2-50-c1) in measurements of ($hk0$) planes at room temperature.  
The observed V$_{2}$O$_{3}$ inclusions are again coherently 
oriented with respect to the CaV$_2$O$_4$ lattice.  The intensities 
of the diffraction peaks related to V$_{2}$O$_{3}$ are similar to those in the 
annealed triarc crystal (an-2-50-c1)  and also vary only slightly upon 
scanning different spots of the crystal which indicates a 
homogeneous distribution of the V$_{2}$O$_{3}$ inclusions with a 
similar volume fraction.  However, the temperature dependence of 
the diffraction pattern is different for the two crystals.  Measurements taken on cooling show that in the 
annealed floating-zone crystal (an-3-074 OFZ), the shape and 
position of the peaks originating from V$_{2}$O$_{3}$ are stable from 
room temperature down to 130~K where the onset of the structural transition occurs.  Around 120~K strong changes are observed similar to the observations around 180~K in the 
annealed triarc crystal (an-2-50-c1).  Below 110~K, the transition 
to the low-temperature monoclinic structure of V$_{2}$O$_{3}$ is 
complete.  Therefore, the temperature for the 
rhombohedral-to-monoclinic transition is reduced by $\sim 60$~K compared to the corresponding temperature in the annealed triarc crystal (an-2-50-c1).  

\section{\label{measure}Magnetization, Magnetic Susceptibility, Heat Capacity
and Thermal Expansion Measurements}

In the following, we describe our results of magnetization, magnetic
susceptibility, heat capacity, and thermal expansion measurements of both
polycrystalline and single crystal samples.  These and additional measurements
consistently identify temperatures at which the antiferromagnetic transition
($T_{\rm N}$), the orthorhombic-to-monoclinic structural transition ($T_{\rm
S}$) and the transition at $\sim 200$~K ($T_{\rm S1}$) occur in these
samples.  In Table~\ref{ordering_temp}, we summarize these transition
temperatures for the different samples obtained using the various  measurements.

\begin{table}
\caption{\label{ordering_temp} Antiferromagnetic ordering (N\'eel) temperature
($T_{\rm N}$), high temperature orthorhombic to low-temperature monoclinic
structural transition temperature ($T_{\rm S}$), and the transition temperature 
at $\sim 200$~K ($T_{\rm S1}$) observed  by static magnetic
susceptibility $\chi$ (peak of d$(\chi T)$/d$T$), heat capacity
$C_{\rm p}$ (peak of $\Delta C_{\rm p}$), thermal expansion
$\alpha$ [peak of $\alpha(T)$, except for $T_{\rm S1}$ where the onset of
$\alpha$ slope change is used], powder synchrotron x-ray diffraction (XRD), 
single crystal neutron diffraction (ND), and single crystal high-energy x-ray diffraction (HEXRD) measurements for polycrystalline
(powder) and single crystal CaV$_2$O$_4$ samples.  The single crystals were
grown using either a triarc furnace or an optical floating zone (OFZ) furnace.}
\begin{ruledtabular}
\begin{tabular}{llcccc}
  Sample & Synthesis & Method & $T_{\rm N}$ (K)& $T_{\rm S}$ (K)  
& $T_{\rm S1}$~(K)\\
\hline
an-2-116 &  powder & $\chi$ & 76 & 147  
&\footnotemark[1]\\
  & 1200~$^\circ$C & $C_{\rm p}$ & 75  & 144 &\footnotemark[1] \\
& & XRD& -- & 150 & 200\\
\hline
an-2-50 & Triarc crystal & $\chi$ & 51 & 108 & \footnotemark[1]\\
&  as-grown &$C_{\rm p}$ & 51 & 108 & \footnotemark[2]\\
& & ND & 53 & 112 & \footnotemark[2]\\
\hline
an-2-50 & Triarc crystal  & $\chi$ & 68 & 133 & 195 \\
  &annealed 1200~$^\circ$C &$C_{\rm p}$ & 68 & 133 & 193\\
  & &$\alpha$ &68 &136 &198\\
& & ND & 69 & 141 & \footnotemark[2] \\
& & HEXRD & -- & 138(2) & 192(7)\footnotemark[3]\\
\hline
an-3-074& OFZ crystal & $\chi$ & 69 & 136 & 192\\
& annealed 1200~$^\circ$C  &$C_{\rm p}$ & 71 & 132 & 191\\
& & ND & 69 & 147 & \footnotemark[2]
\footnotetext[1]{No clear signature of a transition corresponding to  
$T_{\rm S1}$ was observed.}
\footnotetext[2]{Not measured for this sample.}
\footnotetext[3]{The error bar reflects the hysteresis on warming and cooling.}
\end{tabular}
\end{ruledtabular}
\end{table}

\subsection{Magnetization and Magnetic Susceptibility Measurements} 

\begin{figure*}[t]
\includegraphics[width=4in]{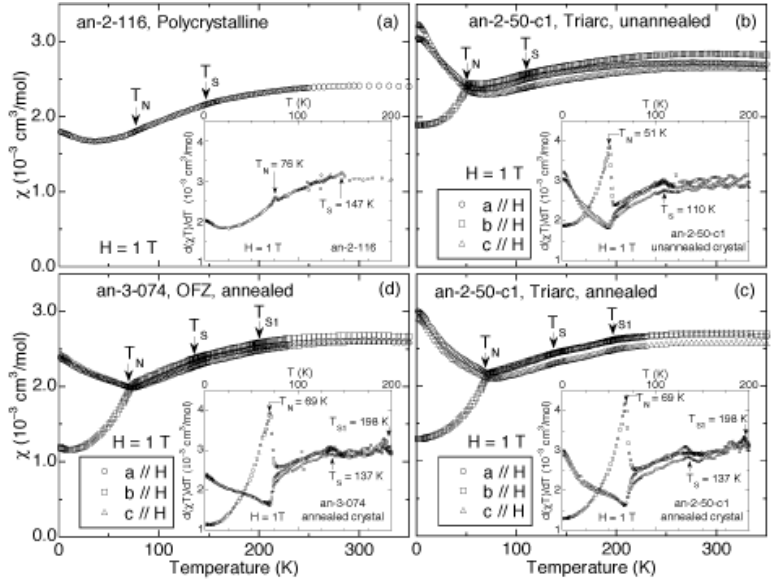}
\caption{Magnetic susceptibility $\chi$ versus temperature in a
field of 1~T of CaV$_2$O$_4$ (a) polycrystalline sample, (b) as-grown
triarc-grown single crystal, (c) annealed triarc-grown single crystal, and (d)
annealed OFZ-grown single crystal.  The axes ($a,\ b,$ or $c$) along which the
measurements were carried out are as indicated.  The insets show d$(\chi
T)$/d$T$ versus $T$ to highlight the transition temperatures.  The oscillatory
behavior of d$(\chi T)$/d$T$ at the higher temperatures, most pronounced in the
inset in (b), is an artifact generated by the SQUID magnetometer.}
\label{chi-1}
\end{figure*}

\begin{figure*}[t]
\includegraphics[width=4in]{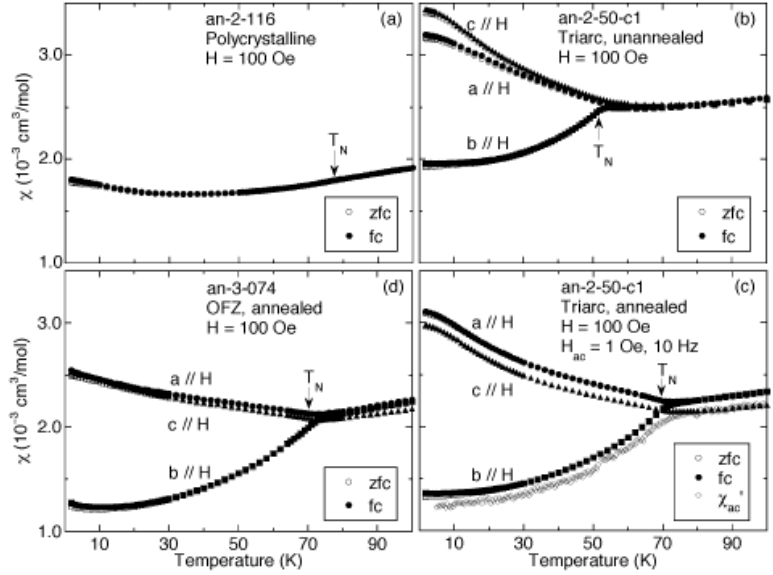}
\caption{Zero-field-cooled and field-cooled susceptibility of
CaV$_2$O$_4$ in a field of 100~Oe measured on (a)
polycrystalline powder, (b) unannealed triarc grown single crystal (c)
annealed triarc grown single crystal, and (d) annealed OFZ grown single
crystal. Part (c) also shows the ac-susceptibility $\chi_{\rm ac}'(T)$
along the easy $b$-axis of the annealed crystal measured in a field
$H_{\rm ac} = 1$~Oe at a frequency of 10~Hz. The antiferromagnetic
transition temperatures $T_{\rm N}$ are marked as shown.}
\label{zfc-fc}
\end{figure*}

\begin{figure}[t]
\includegraphics[width=3in]{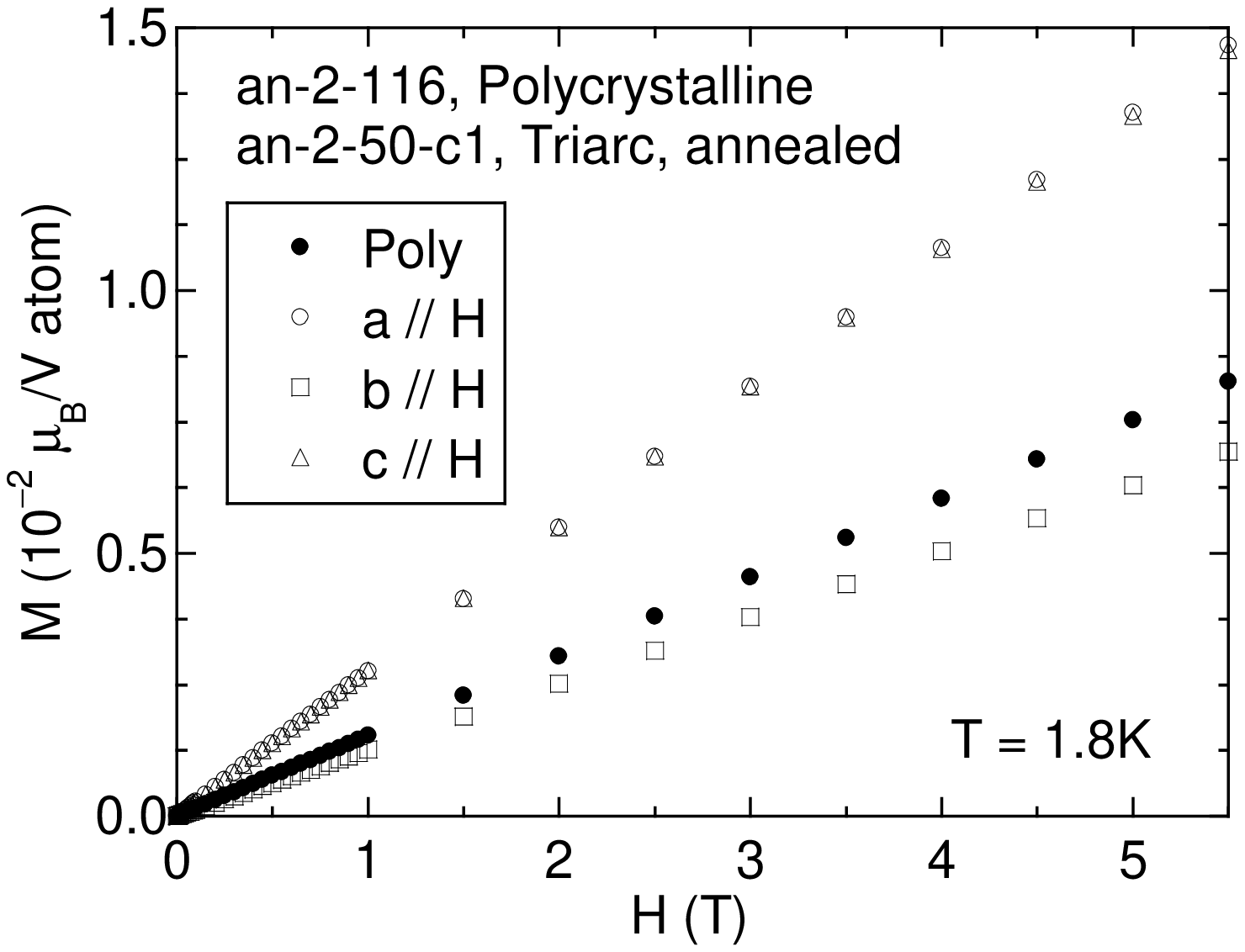}
\caption{Magnetization $M$ versus applied magnetic field $H$ isotherms at a
temperature of 1.8~K of a polycrystalline sample and of an annealed triarc
grown single crystal of CaV$_2$O$_4$.}
\label{M(H)}
\end{figure}

The static magnetic susceptibility versus temperature $\chi(T) \equiv
M(T)/H$ of a polycrystalline sample as well as of the oriented crystals was
measured using a Quantum Design MPMS SQUID magnetometer in a 1~T field
from 1.8~K to 350~K, where $M$ is the magnetization of the sample and $H$ is
the magnitude of the applied magnetic field.  In addition, low field (100~Oe)
zero-field-cooled and field-cooled (zfc, fc) measurements of $M(T)$ at fixed
$H$ were carried out from 1.8~K to 100~K\@.  A Quantum Design MPMS ac SQUID
magnetometer was used to measure the ac susceptibility $\chi_{\rm ac}(T)$ of
the annealed triarc grown crystal from 5 to 100~K in an ac field
$H_{\rm ac} = 1$~Oe and frequency 10~Hz.  The powder was contained in
polycarbonate capsules mounted in clear plastic straws.  Each crystal was
glued to a small piece of clear plastic transparency sheet with GE~7031
varnish or Duco cement, which was then aligned inside the plastic straws
with the $a$, $b$ or $c$ axis direction parallel to the external
magnetic field.   $M(H)$ isotherms were measured in fields up to
$H=5.5$~T at various temperatures. 

The $\chi(T)$ in $H = 1$~T is plotted in
Fig.~\ref{chi-1} for a CaV$_2$O$_4$ polycrystalline sample and for aligned
single crystals grown using a triarc furnace and using an optical floating zone
(OFZ) furnace.  The broad maximum in $\chi(T)$ around 300~K is characteristic
of a low-dimensional spin system with dominant antiferromagnetic exchange
interactions with magnitude of order 300~K\@.  For the single crystal samples, clear evidence is seen for long-range antiferromagnetic ordering at N\'eel temperatures $T_{\rm N} = 51$ to 69~K, depending on the sample.  The easy axis of the antiferromagnetic ordering (with the lowest susceptibility as $T \to 0$) is seen to be the $b$-axis, perpendicular to the zigzag V chains.  At temperatures above $T_{\rm N}$, the susceptibility of the crystals is nearly isotropic, but with small anisotropies which typically showed $\chi_b > \chi_a > \chi_c$.  However, occasionally variations of $\pm 5$\% in the absolute value
of $\chi_i(T)$ were observed between different runs for the same crystal axis
$i$ that we attribute to sample size and positioning effects (radial
off-centering) in the second order gradiometer coils of the Quantum Design MPMS
SQUID magnetometer.\cite{miller,stamenov}

The ordering temperatures observed are marked by vertical
arrows in Fig.~\ref{chi-1} and are highlighted in the plots of $d(\chi T)/dT$
versus  $T$ shown in the insets.  The various transition temperatures are
summarized in Table~\ref{ordering_temp}.  As is typical for a low-dimensional antiferromagnetic system, the polycrystalline sample shows only
a very weak cusp at $T_{\rm N} = 76$~K due to averaging
over the three principal axis directions, but it is still well-defined as observed in the $d(\chi T)/dT$ {\em vs} $T$ plot shown in the inset of Fig.~\ref{chi-1}(a). 
In a related study, $^{17}$O NMR measurements on a polycrystalline sample of
$^{17}$O-enriched CaV$_2$O$_4$ gave a clear signature of antiferromagnetic
ordering at 78~K.\cite{zong}  In contrast to the polycrystalline sample, the
as-grown crystal in Fig.~\ref{chi-1}(b) shows a clear and distinct
antiferromagnetic ordering temperature but with a much lower value $T_{\rm N}
\approx 51$~K\@.  After annealing the
crystals, Figs.~\ref{chi-1}(c) and \ref{chi-1}(d) show that $T_{\rm N}$
increases to $\approx 69$~K, closer to that observed in the polycrystalline
sample.  However, the powder average of the annealed single crystal
susceptibility below $T_{\rm N}$ does not match the susceptibility of the polycrystalline
sample.  The reason for this disagreement is unclear at
this time.  In any case the slow upturn in the susceptibility of the powder
sample below 40~K in Fig.~\ref{chi-1} is evidently intrinsic, due to the powder
average of the anisotropic susceptibilities, and is not due to magnetic
impurities.

The zero-field-cooled (zfc) and field-cooled (fc) $\chi(T)$ measured in a field of 100~Oe for
polycrystalline and single crystal samples of ${\rm CaV_2O_4}$ are plotted in
Figs.~\ref{zfc-fc}(a)--(d).  Also shown in Fig.~\ref{zfc-fc}(c) is
the real part of the ac susceptibility $\chi_{\rm ac}^\prime(T)$ along the easy
$b$-axis direction of the annealed triarc crystal measured in an ac field of
amplitude 1 Oe at a frequency of 10~Hz.   A small irreversibility is observed
in Fig.~\ref{zfc-fc} in all samples between the zfc and fc susceptibilities
below $\sim 30$~K\@.  However, the $\chi_{\rm ac}^\prime(T)$ measurement in
Fig.~\ref{zfc-fc}(c) does not show any peak in that temperature region, ruling
out spin glass-like spin freezing which was suggested to occur in powder
samples from earlier reports.\cite{fukushima, kikuchi}  The slight
irreversibility observed may be associated with antiferromagnetic domain wall
effects.

In Fig.~\ref{M(H)} we show isothermal $M(H)$ measurements up to $H = 5.5$~T
measured at 1.8~K for the polycrystalline sample and for the annealed
triarc-grown single crystal.  The behavior is representative of all samples
measured.  We find that $M$ is proportional to $H$ at fields up to at least 
$\sim 2$~T, indicating the absence of any significant ferromagnetic impurities
and the absence of a ferromagnetic component to the ordered magnetic
structure.

\subsection{Heat Capacity Measurements} 

The heat capacity $C_{\rm p}$ versus temperature $T$ of a sintered
polycrystalline pellet of CaV$_2$O$_4$ as well as of crystals (as-grown and
annealed) was measured using a Quantum Design PPMS system at $T = 1.8$ to
200--300~K in zero applied magnetic field.  The $C_{\rm p}(T)$ was also
measured of a polycrystalline sintered pellet of isostructural (at
room temperature) nonmagnetic CaSc$_2$O$_4$ whose lattice parameters and formula
weight are very similar to those of CaV$_2$O$_4$.\cite{carter} The
CaSc$_2$O$_4$ sample was synthesized from Sc$_2$O$_3$ (99.99\%, Alfa) and
CaCO$_3$ (99.995\%, Aithaca) by reacting a stoichiometric mixture in air at
1000~$^{\circ}$C for 24 hr and then at 1200~$^{\circ}$C for 96 hr with
intermediate grindings, and checked for phase purity using powder XRD.

\begin{figure}[t]
\includegraphics[width=3.3in]{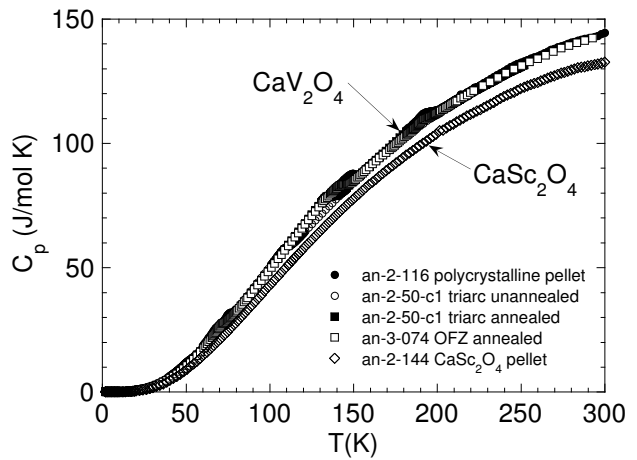}\\
\caption{ Heat Capacity $C_{\rm p}$ versus temperature $T$ in zero magnetic
field of one polycrystalline sample and three single crystal samples of
CaV$_2$O$_4$ and of a polycrystalline sample of isostructural nonmagnetic
CaSc$_2$O$_4$.  On this scale, the data for the four CaV$_2$O$_4$ samples are
hardly distinguishable.}
\label{Cp}
\end{figure}

\begin{figure}[t]
\includegraphics[width=3in]{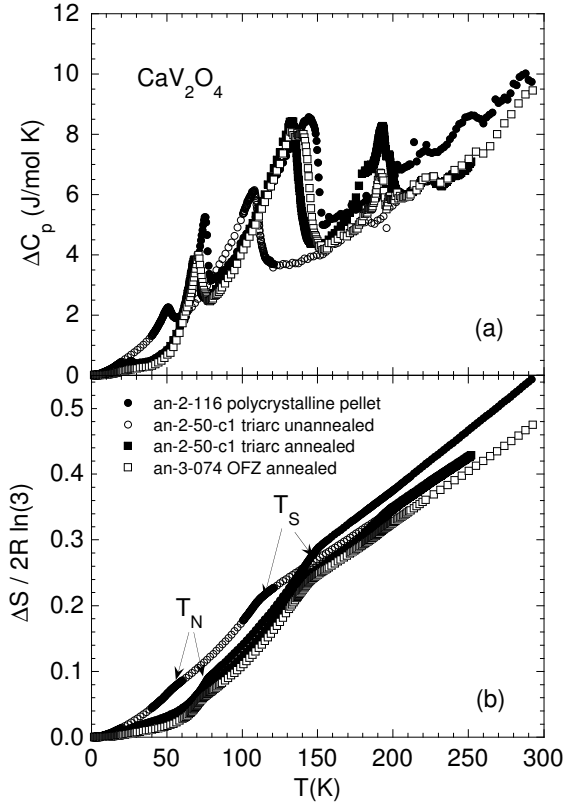}\\
\caption{(a) $\Delta C_{\rm p}$ versus temperature $T$ for four CaV$_2$O$_4$
samples.  Here $\Delta C_{\rm p}$ is the difference between the heat
capacity of CaV$_2$O$_4$ and that of CaSc$_2$O$_4$, but where the
temperature axis of $C_{\rm p}$ for CaSc$_2$O$_4$ was multiplied by 0.9705 to
take account of the difference in the formula weights of CaV$_2$O$_4$ and
CaSc$_2$O$_4$.  (b) Entropy $\Delta S(T)$ associated with the $\Delta C_{\rm
p}(T)$ data in (a), obtained by integrating $\Delta C_{\rm p}/T$ in (a) versus
$T$.  The $\Delta S$ is normalized by the entropy 2R ln(3) of two moles of
disordered spins $S = 1$, where R is the molar gas constant.}
\label{DeltaCp_DeltaS_All}
\end{figure}

\begin{figure}[t]
\includegraphics[width=3in]{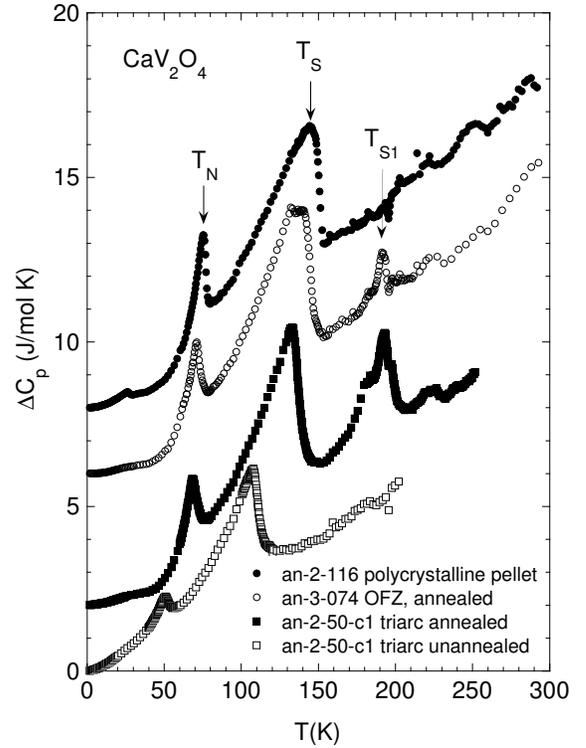}\\
\caption{ $\Delta C_{\rm p}$ versus temperature $T$ for four CaV$_2$O$_4$
samples.  The data are the same as in Fig.~\ref{DeltaCp_DeltaS_All}(a) except
for vertical offsets to separate the data sets.  The symbol $T_{\rm N}$ labels
the long-range antiferromagnetic transition and $T_{\rm S}$ labels the
high temperature orthorhombic to low temperature monoclinic structural
transition. An additional transition at $T_{\rm S1} \sim 200$~K is seen in the two annealed single crystals.  The small anomaly for the AFZ crystal and for the triarc crystal that rides on top of the broader peak appear to be intrinsic to CaV$_2$O$_4$.  The broad peak in the latter crystal appears to be due to coherently grown V$_2$O$_3$ impurity phase that grows during annealing (see text).}
\label{FigCaV2O4_DeltaCp_Shift_All}
\end{figure}

In Fig.~\ref{Cp} we plot the measured $C_{\rm p}(T)$ of four CaV$_2$O$_4$
samples and of isostructural nonmagnetic CaSc$_2$O$_4$.  The difference $\Delta
C_{\rm p}$ versus $T$ for the four CaV$_2$O$_4$ samples is plotted in
Fig.~\ref{DeltaCp_DeltaS_All}(a).  Here $\Delta C_{\rm p}$ is the difference
between the heat capacity of CaV$_2$O$_4$ and that of CaSc$_2$O$_4$, but where
the temperature axis of $C_{\rm p}$ for CaSc$_2$O$_4$ was multiplied by a scaling factor to take account of the difference in the formula weights of CaV$_2$O$_4$ and CaSc$_2$O$_4$.  This factor is given by $[M_{\rm M}$(CaSc$_2$O$_4$)/$M_{\rm M}$(CaV$_2$O$_4)]^{1/2}$=0.9705 where $M_{\rm M}$ is the molar mass of the
respective compound.  If the lattice heat capacity of CaV$_2$O$_4$ and the 
(renormalized) heat capacity of CaSc$_2$O$_4$ had been the same, the difference
$\Delta C_{\rm p}(T)$ would presumably have been the magnetic heat capacity of
CaV$_2$O$_4$.  However, due to the structural transition at $T_{\rm S}$ and the
transition at $T_{\rm S1}$, $\Delta C_{\rm p}(T)$  contains a
lattice contribution as well.  The lattice contribution to $\Delta C_{\rm p}$
is expected to be minimal at low temperatures, where only the long wavelength
acoustic phonon modes are excited, and possibly also above $T_{\rm S1}$ where
the two compounds are known to be isostructural.  

Figure~\ref{DeltaCp_DeltaS_All}(a) shows that the $\Delta C_{\rm p}(T)$
data for the four  CaV$_2$O$_4$ samples are similar except for the different
sizes and temperatures of the anomalies associated with three  transitions.  In
order to more clearly illustrate the differences between samples,
Fig.~\ref{FigCaV2O4_DeltaCp_Shift_All} shows the same data for each sample but
vertically displaced from each other to avoid overlap.  The magnetic ordering
transition at $T_{\rm N}$ as well as the ordering temperatures $T_{\rm S}$ and
$T_{\rm S1}$ are clearly evident from the $\Delta C_{\rm p}(T)$ data in
Fig.~\ref{FigCaV2O4_DeltaCp_Shift_All}.   The ordering temperatures observed are
summarized above in Table~\ref{ordering_temp}.

The entropy versus temperature associated with the $\Delta C_{\rm
p}(T)$ data of each sample in Fig.~\ref{DeltaCp_DeltaS_All}(a) is shown in
Fig.~\ref{DeltaCp_DeltaS_All}(b), obtained from $\Delta S(T) = \int_0^T [\Delta
C_{\rm p}(T)/T] dT$.  In Fig.~\ref{DeltaCp_DeltaS_All}(b), $\Delta S$ is
normalized by the entropy \mbox{2Rln($2S+1$) = 2Rln(3)} for two moles of fully
disordered spins $S = 1$, where R is the molar gas constant.  As noted above, at
least at low temperatures, we associate $\Delta S(T)$ with the magnetic entropy
of the system.  At the antiferromagnetic ordering temperature $T_{\rm N}$, the
normalized value of $\Delta S(T_{\rm N})/2{\rm R}\ln(3) \approx 6$--8\% is very
small and is about the same for all samples.  This small value indicates that
short-range antiferromagnetic ordering is very strong above $T_{\rm N}$ and the
data in Fig.~\ref{DeltaCp_DeltaS_All}(b) indicate that the maximum spin entropy
of the system is not attained even at room temperature.  This is
qualitatively consistent with our estimate $J_1 \approx 230$~K obtained below in
Sec.~\ref{calc} by comparison of our $\chi(T)$ data with calculations of
$\chi(T)$.

The small observed magnetic entropy at $T_{\rm N}$ is consistent with the
values of the heat capacity discontinuities $\Delta C_{\rm AF}$ at $T_{\rm
N}$ in Fig.~\ref{FigCaV2O4_DeltaCp_Shift_All}, as follows.  In mean field
theory, for a system containing $N$ spins $S$ the discontinuity in the
magnetic heat capacity at the ordering temperature for either ferromagnetic or
antiferromagnetic ordering is predicted to be\cite{Smart1966}
\begin{equation}
\Delta C_{\rm AF} =
\frac{5}{2}Nk_{\rm B}\frac{(2S + 1)^2 - 1}{(2S + 1)^2 + 1},
\end{equation} 
where $N$ is the number of spins and $k_{\rm B}$ is Boltzmann's constant.  Using $S =
1$ relevant to V$^{+3}$ and $Nk_{\rm B} = 2$R, where R is the molar gas
constant, one obtains $\Delta C_{\rm AF} = 4$R$ = 33.3$~J/mol~K, where a
``mol'' refers to a mole of CaV$_2$O$_4$ formula units.  From
Fig.~\ref{FigCaV2O4_DeltaCp_Shift_All}, the experimental $\Delta C_{\rm AF}$ is
about 0.5 to 2 J/mol~K, which is only 1.5--6\% of the mean field value.  This
small jump in $C_{\rm p}(T_{\rm N})$ is consistent with the above small value
of $S(T_{\rm N})$.  When short range magnetic ordering removes most of
the magnetic entropy of a system at high temperatures, then thermal effects
associated with three-dimensional magnetic ordering of the system at low
temperatures will necessarily be much smaller than otherwise expected.

\subsection{Thermal Expansion Measurements\label{ThermalExpansion}}  

\begin{figure}[t]
\includegraphics[width=3.3in]{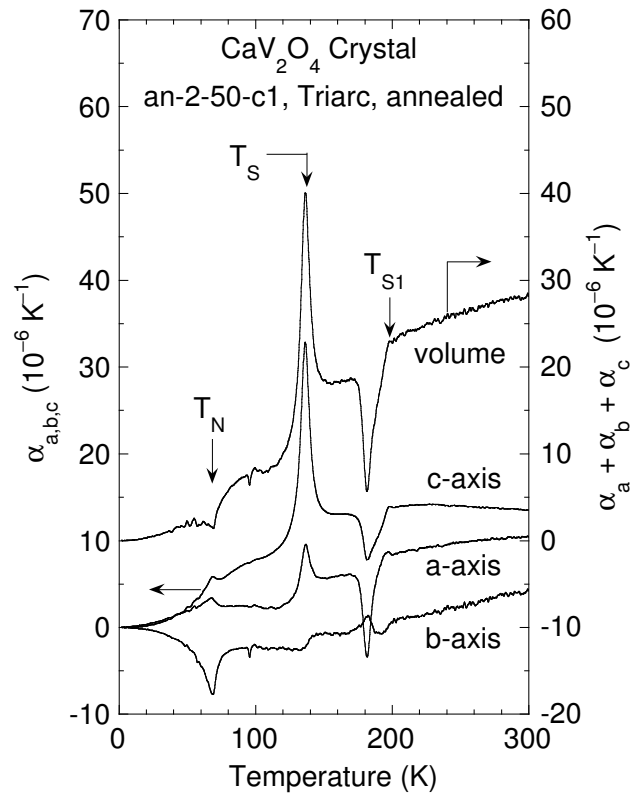}
\caption{Coefficients of linear thermal expansion $\alpha_{a,b,c}$ of an
annealed single crystal of CaV$_2$O$_4$ measured along the orthorhombic $a$,
$b$, and $c$-axes versus temperature (left-hand scale), and the volume expansion
coefficient $\alpha_a + \alpha_b + \alpha_c$ versus temperature (right-hand
scale).}
\label{dil}
\end{figure}

\begin{table}
\caption{\label{LattParChanges} Relative length changes along the orthorhombic
$a$-, $b$- and $c$-axis directions and of the volume $V$ that are associated with the
three transitions at $T_{\rm N} \approx 70$~K, $T_{\rm S} \approx 140$~K, and
$T_{\rm S1}\approx 195$~K in annealed triarc CaV$_2$O$_4$ single crystal
an-2-50-c1.  The temperature $T$ range over which the changes were measured are
as indicated.}
\begin{ruledtabular}
\begin{tabular}{lccc}
 & $T_{\rm N}$ & $T_{\rm S}$ & $T_{\rm S1}$ \\
$T$ range (K) & 52--75 & 120--153 & 174--198 \\
\hline
$\Delta a/a\ (10^{-5})$ & 1.2 & 4.9 & $-$7.5 \\
$\Delta b/b\ (10^{-5})$ & $-$4.3 & $-0.9$ & $-$0.1 \\
$\Delta c/c\ (10^{-5})$ & 1.1 & 19.6 & $-$6.3 \\
$\Delta V/V\ (10^{-5})$ & $-$2.0 & 23.6 & $-$13.9 \\
\end{tabular}
\end{ruledtabular}
\end{table}

The thermal expansion of the annealed triarc grown CaV$_2$O$_4$ crystal
an-2-50-c1 was measured versus temperature using capacitance
dilatometry\cite{dil-RSI} from 1.8 to 300~K along the three orthorhombic axes
$a$, $b$ and $c$.  The crystal is  the same annealed triarc crystal measured by
magnetic susceptibility and heat capacity in Figs.~\ref{chi-1} and
\ref{zfc-fc}(c) and in Figs.~\ref{Cp}--\ref{FigCaV2O4_DeltaCp_Shift_All},
respectively.  In Fig.~\ref{dil} the linear coefficients of thermal expansion
are plotted versus temperature (left-hand scale), along with the volume thermal
expansion coefficient (right-hand scale).  At high temperatures $T \sim 300$~K
the $\alpha$ values tend to become temperature independent.  Below 200~K, the
ordering transitions observed above in the magnetic susceptibility and heat
capacity are reflected in distinct anomalies in the thermal expansion
coefficients at the corresponding temperatures.  The ordering temperatures
observed are summarized above in Table~\ref{ordering_temp}.  

The normalized length changes along the orthorhombic $a,\ b$ and $c$ axis directions
and the normalized change in the volume $V$ associated with the three
transitions at $T_{\rm N}$, $T_{\rm S}$ and $T_{\rm S1}$ are listed in
Table~\ref{LattParChanges}.  These changes were calculated by determining the
areas under the respective peaks in the thermal expansion coefficients in
Fig.~\ref{dil}, and then subtracting the estimated respective background
changes over the same temperature intervals.

\section{\label{calc} Analysis of Experimental Data}

\subsection{Origin of the Transition at $\bm{T_{\rm S1} \sim 200}$~K in Annealed CaV$_{\bm 2}$O$_{\bm 4}$ Single Crystals}

From Table~\ref{LattParChanges} above, the relative volume change of the annealed triarc crystal an-2-50-c1 on heating through $T_{\rm{S1}}$ from the thermal expansion data is $\Delta V/V \approx -1.4 \times 10^{-4}$.  This value is about 1\% of the value for pure V$_2$O$_3$ at its transition.\cite{Whan70} The height of the heat capacity anomaly above ``background'' in Fig.~\ref{FigCaV2O4_DeltaCp_Shift_All} for this crystal is $\approx 2.5$~J/mol~K, which is about 0.8\% of the value\cite{Keer76} at the structural transition for pure V$_2$O$_3$.  These estimates are both consistent with our estimates from x-ray diffraction data in Sec.~\ref{struct} of a 1--2 percent volume fraction of V$_2$O$_3$ in this crystal.  Our data therefore indicate that for the  annealed triarc-grown crystal (an-2-50-c1), the anomalous features found above in the heat capacity and thermal expansion data at $T_{\rm{S1}}$ arise mainly from this transition in the V$_2$O$_3$ impurity phase.

Furthermore, the temperature
dependences of the linear thermal expansion coefficients at the transition 
$T_{\rm S1}\approx 200$~K in Fig.~\ref{dil} are significantly different than near the transitions $T_{\rm N} \approx 70$~K and $T_{\rm S} \approx 140$~K.  There appears to be a discontinuity in the slopes of $\alpha_i(T)$ as the transition
$T_{\rm S1}$ is approached from above, whereas a continuous change in the
slopes occurs as $T_{\rm S}$ and $T_{\rm N}$ are approached from above.  The reason for this difference is evidently that the former transition is mainly due to the \emph{first order} structural transition in the V$_2$O$_3$ coherently-grown impurity phase in this annealed crystal as investigated previously in Sec.~\ref{AnnTriarc}, whereas the latter two transitions are \emph{second order}.

However, we also found in Sec.~\ref{struct} that for the annealed floating-zone crystal (an-3-074 OFZ), the structural transition of the V$_2$O$_3$ impurity phase was reduced by $\sim 60$~K from that of the V$_2$O$_3$ impurity phase in the annealed triarc grown crystal, and hence cannot be responsible for heat capacity anomaly at $T_{\rm{S1}} \sim 200$~K for the annealed float-zone crystal in Fig.~\ref{FigCaV2O4_DeltaCp_Shift_All}.  Indeed, the relatively small heat capacity anomaly at $T_{\rm{S1}}$ in Fig.~\ref{FigCaV2O4_DeltaCp_Shift_All} for the float-zone crystal appears to also be present at the same temperature for the annealed triarc-grown crystal, but rides on top of a broader anomaly that is evidently due to the structural transition of the V$_2$O$_3$ impurity phase in that crystal.  Furthermore, the double peak structure in the heat capacity for the annealed triarc crystal at $T_{\rm S} \approx 140$~K evidently arises due to the overlap of the onsets of the structural transitions in V$_2$O$_3$ and CaV$_2$O$_4$.

An issue of interest is the cause(s) of the variability in the structural rhombohedral-to-monoclinic transition temperature of the coherently grown V$_2$O$_3$ impurity phase in our annealed CaV$_2$O$_4$ crystals.  Due to the first order nature of the transition, the transition is hysteretic.  The transition temperature of bulk stoichiometric V$_2$O$_3$ has been reported to be at $\approx 170$~K on heating and $\approx 150$~K on cooling.\cite{Keer76,Moon1970,Menth1970,Shivashankar1983}  The transition temperature \emph{decreases} rapidly under pressure.\cite{Carter1994}  A pressure of only 9~kbar lowers the transition temperature by 60~K, and the transition is completely suppressed at a pressure of $\approx 20$~kbar.\cite{Carter1994}  The transition temperature is also rapidly \emph{suppressed} if the sample contains V vacancies; a crystal of composition V$_{1.985}$O$_3$ showed a transition temperature of $\approx 50$~K\@.\cite{Shivashankar1983}  Furthermore, it was found that when V$_2$O$_3$ is epitaxially grown on LiTaO$_3$, the transition temperature is \emph{enhanced} from the bulk value by 20 K.\cite{Yonezawa04} Given the possibilities of compressive or tensile forces acting on the V$_2$O$_3$ due to the epitaxial relationship of the V$_2$O$_3$ impurity with the CaV$_2$O$_4$ host and the possibility of nonstoichiometry of the V$_2$O$_3$ impurity phase, one can see how the transition temperature of the V$_2$O$_3$ might be \emph{depressed or enhanced} from the bulk value by $\approx 30$~K as we found for the coherently grown V$_2$O$_3$ impurity phases in our two annealed crystals in Sec.~\ref{HEXRD}.  

In summary, then, it appears that there is an intrinsic phase transition in the two annealed CaV$_2$O$_4$ crystals at about 200~K that has no obvious source.  We speculate that this transition may be the long-sought chiral phase transition originally postulated by Villain,\cite{Villain1977} where there is long-range chiral order but no long-range spin order below the transition temperature, and the long-range chiral order is lost above the transition temperature.

\subsection{Magnetic Susceptibility and Magnetic Heat Capacity}

In separate experiments to be described elsewhere,\cite{Pieper2008} we have
carried out inelastic neutron scattering measurements of the magnetic excitation
dispersion relations for ${\rm CaV_2O_4}$ single crystals.  We find that the
dispersion along the $c$-axis (in the vanadium chain direction) is significantly larger
than in the two perpendicular directions.  Above the N\'eel temperature $T_{\rm
N}$, the magnetic susceptibility in Fig.~\ref{chi-1} is nearly isotropic.  
Thus a quasi-one-dimensional Heisenberg model appears to be appropriate for the
spin interactions in ${\rm CaV_2O_4}$.  

The crystal structure suggests the presence of spin $S = 1$ zigzag spin
chains along the orthorhombic $c$-axis.  We  report here exact diagonalization
(ED) calculations of the magnetic spin susceptibility versus temperature
$\chi(T)$ and the magnetic heat capacity $C(T)$ of spin $S = 1$ $J_1$-$J_2$
Heisenberg chains containing $N = 8$, 10, and 12 spins for $J_2/J_1$ ratios
from $-1$ to 5, and containing 14 spins for $J_2/J_1 = 0$.  We also report the
results of quantum Monte Carlo (QMC) simulations of $\chi(T)$ and $C(T)$.  These
simulations were carried out with  the ALPS directed loop
application\cite{ALPS} in the stochastic series expansion
framework\cite{Sandvik} for chains with $N = 30$ and 60 spins and $J_2/J_1 = 0$.  Here
$J_1$ and $J_2$ are the nearest-neighbor and next-nearest-neighbor interactions
on a linear chain, respectively.  The spin Hamiltonian is the $\lambda=1$
special case of Eq.~(\ref{EqH1}), given by
\begin{equation}
{\cal H} = \sum_{i=1}^N (J_1 \mathbf{S}_i\cdot \mathbf{S}_{i+1} + J_2
\mathbf{S}_i\cdot \mathbf{S}_{i+2}),
\end{equation}
where $\mathbf{S}$ is a spin-1 operator.  Periodic boundary conditions are
imposed, so the chains become rings.  $J_1$ is always positive
(antiferromagnetic) whereas $J_2$ was taken to be either positive or negative
(ferromagnetic).  This chain is topologically the same as a zigzag chain in
which $J_1$ is the nearest-neighbor interaction between the two legs of the
zigzag chain and $J_2$ is the nearest-neighbor interaction along either leg of
the zigzag chain.  For $J_2 = 0$, the $N$ spins are all part of the same
nearest-neighbor exchange ($J_1$) chain.  For $J_1 = 0$, two independent
isolated equivalent chains are formed, each containing $N/2$ spins and with
nearest-neighbor exchange $J_2$.  This effect can be quantified using the $T=0$
correlation length $\xi$ which has been computed in Ref.~\onlinecite{KRS96}. 
We find that we can reach ratios $N/\xi$ which are at least 2 for $J_2/J_1
\lesssim 0.6$ whereas $\xi$ becomes comparable to or even bigger than the
system sizes $N$ which are accessible by ED for larger $J_2/J_1$.  Accordingly,
our finite chain data become a poorer approximation to the infinite $J_1$-$J_2$
chain for large
$J_2/J_1$.  This is exemplified below in Fig.~\ref{FigJ1J2CaV2O4ChiMax_Tmax}
where the data for chains containing different numbers $N$ of spins exhibit an
increasing divergence from each other with increasing $J_2/J_1$.

We will compare the spin susceptibility calculations with the experimental
magnetic susceptibility data to estimate the $J_1$ and $J_2/J_1$ values in the
$J_1$-$J_2$ chain model for ${\rm CaV_2O_4}$.  These values will also be used as
input to compare the calculated magnetic heat capacity versus temperature with
the experimental heat capacity data.

\subsubsection{Magnetic Susceptibility}

\begin{figure}[t]
\includegraphics[width=3in]{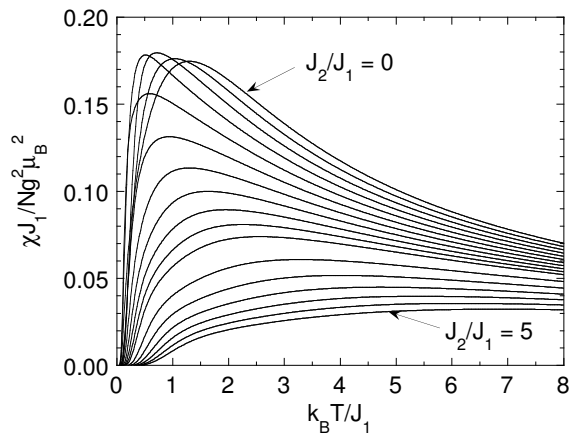}
\caption{Calculated magnetic spin susceptibility $\chi$ for spin $S = 1$ 
$J_1$-$J_2$ Heisenberg chains containing $N = 12$ spins versus
temperature $T$, where $J_1$ and $J_2$ are the
nearest-neighbor and next-nearest-neighbor exchange interactions in the
chain.  The curves from top to bottom on the right are for $J_2/J_1 = 0$, 0.2, 0.4, \ldots, 2.0, 2.5, \ldots, 5.0.}
\label{FigN_12ChiJ1b}
\end{figure}

The calculated  magnetic spin susceptibility $\chi(T)$ data for the spin $S = 1$
$J_1$-$J_2$ Heisenberg chain model are in the dimensionless form 
\begin{equation}
\frac{\chi J_1}{N g^2 \mu_{\rm B}^2}\ \ {\rm versus}\ \ \frac{k_{\rm B}
T}{J_1}, 
\label{Eqchi(T)}
\end{equation} 
where $N$ is the number of spins, $g$ is the spectroscopic splitting factor
($g$-factor) of the magnetic moments for a particular direction of the
applied magnetic field with respect to the crystal axes, $\mu_{\rm B}$ is the
Bohr magneton, and $k_{\rm B}$ is Boltzmann's constant.  Calculated
$\chi(T)$ data sets for $N = 12$ were obtained by exact diagonalization
assuming periodic boundary conditions (ring geometry) for $J_2/J_1$ ratios of
$-1,\ -0.8,$ ..., 2.0, 2.5, ..., 5.  Examples of the calculations for a
selection of $J_2/J_1$ values are shown in Fig.~\ref{FigN_12ChiJ1b}.  Each
chain has an energy gap (spin gap) from the nonmagnetic singlet ground state to
the lowest magnetic excited states.\cite{KRS96}   No interchain (between
adjacent zigzag chains) interactions are included in the calculations.  These
calculations are not expected to apply to our system at low temperatures where
we see long-range antiferromagnetic ordering.  However, we expect to be
able to obtain approximate estimates of $J_1$ and $J_2$ by fitting the observed
susceptibility data around the broad peak in the susceptibility at $\approx
300$~K\@.

\begin{figure}[t]
\includegraphics[width=3in]{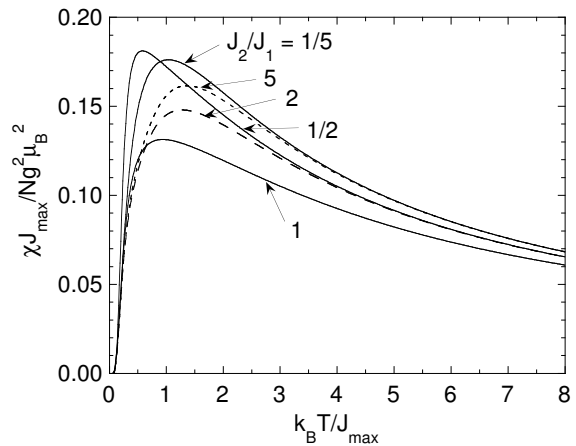}
\caption{Magnetic susceptibility $\chi$ versus temperature $T$ for the spin $S =
1$ $J_1$-$J_2$ Heisenberg chain containing $N = 12$ spins.  Here, $J_{\rm max}$
= max($J_1,J_2)$.  Pairs of curves for $J_2/J_1$ ratios that are reciprocals of
each other become the same at high temperatures.}
\label{FigJ1J2ChainN_12ChiJmax}
\end{figure}

At high temperatures $k_{\rm B}T \gg J_{\rm max}$, where $J_{\rm max} =$
max($J_1,J_2$), one expects that the calculated $\chi J_{\rm max}$ versus
$k_{\rm B}T/J_{\rm max}$ should be nearly the same upon interchange of $J_1$ and
$J_2$, i.e., the same for pairs of $J_2/J_1$ ratios that are reciprocals of each
other.  [This is because all spins in the zigzag chain are equivalent, and at high temperatures the Curie-Weiss law is obtained.  The Weiss temperature $\theta$ only depends on the numbers of nearest neighbors $z$ to a given spin and the corresponding interaction strengths $J$ ($\theta \sim z_1J_1+z_2J_2$ with $z_1=z_2=2$), which is invariant upon interchange of $J_1$ and $J_2$.] This expectation is confirmed in Fig.~\ref{FigJ1J2ChainN_12ChiJmax} where such
plots are shown for $J_2/J_1 =$ 1/5 and 5; 1/2 and 2; and 1.  The
data for $J_2/J_1 = 1/2$ and~2, and for $J_2/J_1 = 1/5$ and~5,  are seen to
be about the same for temperatures $k_{\rm B}T/J_{\rm max} \gtrsim 4$,
respectively.

\begin{figure}[t]
\includegraphics[width=3in]{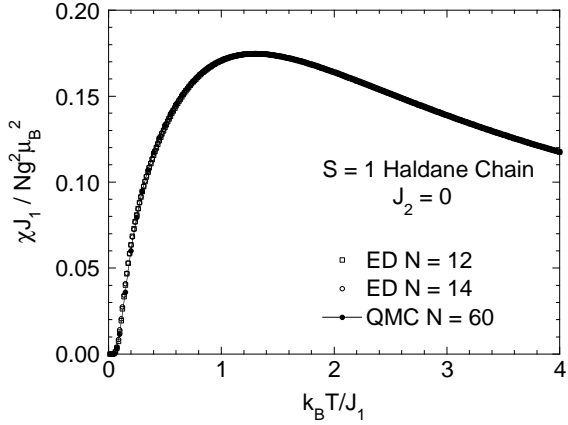}
\caption{Magnetic susceptibility $\chi$ versus temperature $T$ calculations for
the spin $S = 1$ Heisenberg chain with nearest neighbor exchange interaction
$J_1$ and next-nearest-neighbor interaction $J_2 = 0$.  The calculations were
carried out using exact diagonalization (ED) for $N = 12$ and 14, and by
quantum Monte Carlo (QMC) for $N = 60$.}
\label{FigJ2=0_N=12,14,60}
\end{figure}

\begin{figure}[t]
\includegraphics[width=3in]{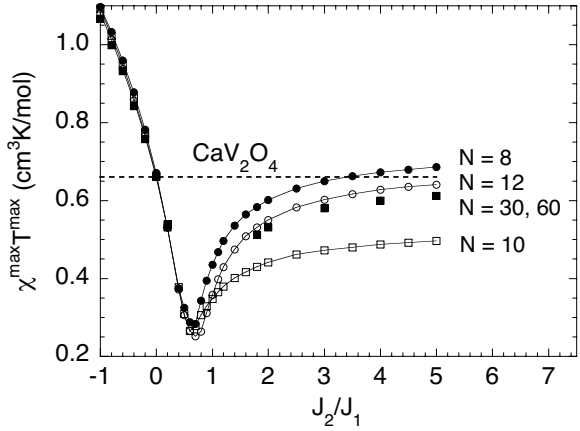}
\caption{The product $\chi^{\rm max}T^{\rm max}$ versus $J_2/J_1$ for the
spin $S = 1$ $J_1$-$J_2$ Heisenberg chain containing $N = 8$, 10, 12, 30 or 60
spins, where we have assumed $N = 2N_{\rm A}$ and $g = 1.97$.  The data for $N
= 8$, 10, and 12 were obtained using exact diagonalization calculations.  The
data for $N = 30$ and~60 were obtained from quantum Monte Carlo simulations; most of these data were obtained for $N = 60$, except for $J_2/J_1 = 0.2$, 1.8 and 2 where we used $N = 30$.  The horizontal dashed line is the
experimental value of $\chi^{\rm max}T^{\rm max}$ for ${\rm CaV_2O_4}$ from
Eq.~(\protect\ref{EqChimaxTmaxExp}).  Comparison of this experimental value
with the calculations indicates that within the $J_1$-$J_2$ model, $J_2/J_1
\approx 0$  (or $J_1/J_2 \approx 0$)  in ${\rm CaV_2O_4}$.}
\label{FigJ1J2CaV2O4ChiMax_Tmax}
\end{figure}

The experimental magnetic susceptibility data of CaV$_2$O$_4$ will be fitted below by the calculated susceptibility $\chi(T)$ of a single $S=1$ chain $(J_2/J_1=0)$.  Such integer-spin chains are known as Haldane chains.
\cite{Haldane}  We will therefore test here the sensitivity of the calculations
to the number of spins $N$ in the chain for this fixed $J_2/J_1$ value.  Shown
in Fig.~\ref{FigJ2=0_N=12,14,60} are exact diagonalization (ED) calculations of
$\chi(T)$ for $J_2/J_1 = 0$ and $N = 12$ and 14, and quantum Monte Carlo (QMC)
simulations for $J_2/J_1 = 0$ and $N = 60$.  On the scale of the figure, the
results of the three calculations can hardly be distinguished.  These data are
fully consistent with previous transfer-matrix renormalization-group results
for $\chi(T)$.\cite{Xiang, FW05}  In Table~\ref{TableJ2=0ChainChiCmax}, the
values of the maxima in the magnetic susceptibility $\chi^{\rm max}$ and also
of the magnetic heat capacity $C^{\rm max}$  (see below) and the  temperatures
$T_\chi^{\rm max}$ and $T_C^{\rm max}$ at which they respectively occur are
listed for the different calculations.  For all three calculations, the maximum
in the susceptibility occurs at about the same temperature $k_{\rm B}T^{\rm
max}_\chi/J_1 \approx 1.30$, which may be compared with previous values of
1.35  (Ref.~\onlinecite{deJongh2001}) and 1.32(3).\cite{Pedrini2004}  Probably
the most accurate values for the susceptibility are those of
Ref.~\onlinecite{FW05}, as listed in Table~\ref{TableJ2=0ChainChiCmax}.

From the theoretical $\chi(T)$ data, for each value of $J_2/J_1$ one can obtain
the value of the normalized temperature $k_{\rm B} T^{\rm max}/J_1$ at which the
maximum in the susceptibility occurs, and the normalized value of the
susceptibility $\chi^{\rm max} J_1/N g^2
\mu_{\rm B}^2$ at the maximum.  For a given value of $J_2/J_1$, the product of
these two values is a particular dimensionless number
\begin{equation} {\chi^{\rm max}  T^{\rm max}\over N g^2 \mu_{\rm
B}^2/k_{\rm B} }
\label{chimaxTmax}
\end{equation} 
that does not contain either exchange constant.  

The spectroscopic splitting tensor ($g$-tensor) for vanadium cations is found
to not depend much on either the  oxidation (spin) state of the V cation or on
its detailed environment in insulating hosts.  The physical origin of this
insensitivity is the small magnitude of the spin-orbit coupling constant for
the vanadium atom.  Typical values for the spherically-averaged $g$-factor
$\langle g\rangle$ are between approximately 1.93 and 1.97, with the individual
components of the diagonal $g$-tensor lying between 1.90 and 2.00.  For example,
for V$^{+2}$ in single crystals of AgCl, one obtains $\langle g\rangle =
1.970(3)$;\cite{Shields1970} for V$^{+3}$ in guanidinium vanadium sulfate
hexahydrate, $\langle g\rangle = 1.94(1)$;\cite{Ashkin1978} for V$^{+4}$ in
TiO$_2$, $\langle g\rangle = 1.973(4)$.\cite{Kubec1972}  

On the basis of the above discussion we set $g = 1.97$ for the V$^{+3}$ spin $S
= 1$ in Eq.~(\ref{chimaxTmax}).  Then setting $N = 2N_{\rm A}$, where $N_{\rm
A}$ is Avogadro's number and the factor of 2 comes from two atoms of V per
formula unit of CaV$_2$O$_4$, the expression in Eq.~(\ref{chimaxTmax}) becomes
\begin{equation} 
{\chi^{\rm max}  T^{\rm max}\over 2.91\ {\rm cm^3\,K/mol} }\ ,
\label{EqChimaxTmax}
\end{equation} 
where a ``mol'' refers to a mole of CaV$_2$O$_4$ formula units.  Then from
Eq.~(\ref{EqChimaxTmax}) and the calculated $\chi(T)$ data for different values
of $J_2/J_1$, the calculated  $\chi^{\rm max}  T^{\rm max}$ versus $J_2/J_1$ for
CaV$_2$O$_4$ is shown in Fig.~\ref{FigJ1J2CaV2O4ChiMax_Tmax}.  From the figure,
the dependence of $\chi^{\rm max} T^{\rm max}$ on  $J_2/J_1$ is about the same 
for $N=8$, $10$, and $12$ for $J_2/J_1 \lesssim 0.6$
which is consistent with a short correlation length $\xi \lesssim 6$
for $0 \le J_2/J_1 \le 0.6$ (see Ref.~\onlinecite{KRS96}). However, the
curves for the different values of $N$ are quite different at larger
values of $J_2/J_1$; the behavior versus $N$ even becomes nonmonotonic
in this parameter region. The QMC results also shown in
Fig.~\ref{FigJ1J2CaV2O4ChiMax_Tmax} nevertheless indicate that the ED
calculations for $N=12$ sites yield a good approximation  to the infinite $N$
limit also for $J_2/J_1 \ge 1.8$.  Unfortunately, the QMC sign problems are so
severe in the region $0.2 < J_2/J_1 < 1.8$ that here we cannot resolve the maximum
of $\chi$ with our QMC simulations.

\begin{table}
\caption{\label{TableJ2=0ChainChiCmax} Calculated values of the maxima in the
magnetic spin susceptibility $\chi^{\rm max}$ and magnetic heat capacity $C^{\rm
max}$ and temperatures $T_\chi^{\rm max}$ and $T_C^{\rm max}$ at which they
occur, respectively, for the linear spin $S = 1$ Heisenberg chain (Haldane
chain) with nearest-neighbor exchange interaction $J_1$ and
next-nearest-neighbor interaction $J_2 = 0$.  The results of exact
diagonalization (ED) and quantum Monte Carlo (QMC) calculations are shown. 
Here $N$ is the number of spins in the chain, $g$ is the $g$-factor, 
$\mu_{\rm B}$ is the Bohr magneton, and $k_{\rm B}$ is Boltzmann's constant.
Also included are the results of Ref.~\onlinecite{FW05}, which are probably the
most accurate values currently available for the susceptibility.}
\begin{ruledtabular}
\begin{tabular}{lllll}
 & $\frac{\chi^{\rm max}J_1}{Ng^2\mu_{\rm B}^2}$ & $\frac{k_{\rm B}T_\chi^{\rm
max}}{J_1}$ & $\frac{C^{\rm max}}{Nk_{\rm B}}$ & $\frac{k_{\rm B}T_C^{\rm
max}}{J_1}$
\\
\hline
QMC $N=60$ & 0.174686(9) & 1.301(10) & 0.5431(4) & 0.857(10) \\
ED $N = 12$ & 0.174662 & 1.2992 & 0.5520 & 0.8295 \\
ED $N = 14$ & 0.174677 & 1.2980 & 0.5466 & 0.8398 \\
Ref.~\onlinecite{FW05} & 0.17496(2) & 1.2952(16) && \\
\end{tabular}
\end{ruledtabular}
\end{table}

\begin{figure}[t]
\includegraphics[width=3in]{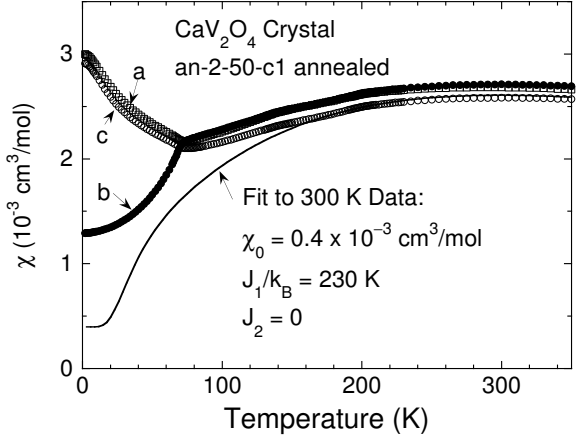}
\caption{Calculated magnetic susceptibility $\chi$ versus temperature $T$ 
for the $J_1$-$J_2$ chain with $J_1/k_{\rm B} = 230$~K, $J_2 = 0$, and $g =
1.97$, and with a temperature-independent orbital contribution $\chi_0 = 0.4
\times 10^{-3}$~cm$^3$/mol (solid curve).  The calculated
spin susceptibility at $T = 0$ is zero, so the zero-temperature
value of the calculated solid curve is $\chi_0$.  Experimental data for annealed
CaV$_2$O$_4$ crystal 2-50-c1 from Fig.~\protect\ref{chi-1} are also shown. 
Comparison of these data for $T\to 0$ with the calculated curve shows that the
spin susceptibility along the easy $b$-axis of CaV$_2$O$_4$ is rather large
for $T \to 0$.}
\label{Figan-2-50-c1_Chi_Fit}
\end{figure}

The experimental susceptibility $\chi_{\rm exp}(T)$ data for
CaV$_2$O$_4$ in Fig.~\ref{chi-1} can be written as the sum $\chi_{\rm exp}(T) =
\chi(T) + \chi_0$, where $\chi(T)$ is the spin susceptibility (which is the
part calculated above) and $\chi_0$ is the temperature-independent orbital
susceptibility.  From the data in Ref.~\onlinecite{jones} for V$_2$O$_3$, we
estimate $\chi_0 \sim 0.4 \times 10^{-3}\ {\rm cm^3/mol}$ for CaV$_2$O$_4$. 
From Fig.~\ref{chi-1} we then obtain the experimental value for the \emph{spin}
susceptibility at the maximum  $\chi^{\rm max} \approx 2.2 \times 10^{-3}\ {\rm
cm^3/mol}$ and for the temperature at the maximum $T_\chi^{\rm max} \approx
300$~K, yielding for CaV$_2$O$_4$
\begin{equation}
\chi^{\rm max}T^{\rm max} \approx 0.66\ {\rm cm^3\ K/mol}\ .
\label{EqChimaxTmaxExp}
\end{equation} 
Comparison of this value with the theoretical spin susceptibility data in
Fig.~\ref{FigJ1J2CaV2O4ChiMax_Tmax} yields $J_2/J_1
\approx 0$ (or $J_1/J_2 \approx 0$).  This ratio of $J_2/J_1$ is quite  different from
the value of unity that we and others\cite{fukushima, kikuchi} initially
expected.  The temperature $T_\chi^{\rm max}\approx 300$~K,
combined with $k_{\rm B}T_\chi^{\rm max}/J_1 \approx 1.30$ from
Table~\ref{TableJ2=0ChainChiCmax}, yields $J_1/k_{\rm B} = 230$~K\@.   Although
the numerical results shown in  Fig.~\ref{FigJ1J2CaV2O4ChiMax_Tmax} are the
least accurate in the vicinity of $J_2/J_1 \approx 1$, it seems rather unlikely
that the value of $\chi^{\rm max} T^{\rm max}$ obtained from the $J_1$-$J_2$
chain model in the region $0.6 \lesssim J_2/J_1 \lesssim 1.8$ could be consistent
with the value in Eq.~(\ref{EqChimaxTmaxExp}) expected for CaV$_2$O$_4$.

The
calculated total susceptibility versus temperature for $J_1/k_{\rm B} = 230$~K,
$J_2 = 0$ and $\chi_0 = 0.4 \times 10^{-3}$~cm$^3$/mol is shown in
Fig.~\ref{Figan-2-50-c1_Chi_Fit}.   Also shown are the experimental
susceptibility data for annealed CaV$_2$O$_4$ crystal an-2-50-c1 from
Fig.~\ref{chi-1}, where an excellent fit of the average anisotropic $\chi(T)$
data near 300~K is seen.  

Above the N\'eel temperature, one sees from Figs.~\ref{chi-1} and
\ref{Figan-2-50-c1_Chi_Fit} that the susceptibility is nearly isotropic.  The
relatively small anisotropy observed can come from anisotropy in the orbital Van
Vleck paramagnetic susceptibility, from $g$-anisotropy arising from spin-orbit
interactions, from single-ion anisotropy of the form $DS_z^2 + E(S_x^2 -
S_y^2)$, and/or from anisotropy in the spin exchange part of the spin
Hamiltonian.  The relative importances of these sources to the observed
susceptibility anisotropies are not yet clear.  The experimental data below
200~K in Fig.~\ref{Figan-2-50-c1_Chi_Fit} increasingly deviate from the fit
with decreasing temperature.  This suggests that other interactions besides
$J_1$ and $J_2$ and/or the presence of magnetic anisotropies may be important to
determining the spin susceptibility above $T_{\rm N}$ in CaV$_2$O$_4$. 

For collinear antiferromagnetic (AF) ordering, one nominally expects the spin
susceptibility along the easy axis to go to zero as $T\to 0$.  Comparison of
the theoretical curve with the experimental easy axis ($b$-axis) data
$\chi_b(T)$ in Fig.~\ref{Figan-2-50-c1_Chi_Fit} indicates that the
zero-temperature $b$-axis spin susceptibility is not zero, but is instead a
rather large value $\chi^{\rm spin}_b(T\to 0) \approx 0.9 \times
10^{-3}$~cm$^3$/mol.  This finite spin susceptibility indicates either that the
spin structure in the AF state is not collinear, that not all vanadium spins
become part of the ordered magnetic structure below $T_{\rm N}$, and/or that
quantum fluctuations are present that induce a nonzero spin susceptibility. 
Such quantum fluctuations can arise from the low-dimensionality of the spin
lattice and/or from frustration effects.  As discussed in the Introduction, our
recent NMR and magnetic neutron diffraction experiments on single crystal CaV$_2$O$_4$ indicated that the magnetic structure at 4~K is noncollinear,\cite{zong,Pieper2008B} which can at least
partially explain the nonzero spin susceptibility along the (average) easy
$b$-axis at low temperatures.  In addition, the reduction in the local ordered
moment 1.0--$1.6~\mu_{\rm B}$/(V atom) of the \emph{ordered} vanadium spins found
in these studies from the expected value $gS\mu_{\rm B} = 2~\mu_{\rm
B}$/(V atom) suggests that quantum zero-point spin fluctuations could be strong and could contribute to the large finite $\chi_b^{\rm spin}(T\to 0)$.

\subsubsection{Magnetic Heat Capacity}

\begin{figure}[t]
\includegraphics[width=3in]{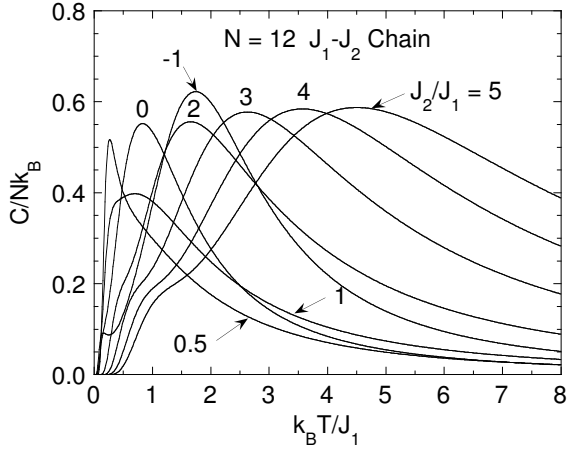}
\caption{Magnetic heat capacity $C$ versus temperature $T$ for the spin $S = 1$ 
$J_1$-$J_2$ Heisenberg chain with $J_2/J_1$ values from $-1$ to 5, calculated
using exact diagonalization with $N = 12$.  Here $N$ is the number of spins,
$k_{\rm B}$ is Boltzmann's constant, and $J_1 > 0$ and $J_2$ are the
nearest-neighbor and next-nearest-neighbor exchange interactions, respectively.}
\label{FigJ1J2ChainN_12Cmag}
\end{figure}

\begin{figure}[t]
\includegraphics[width=3in]{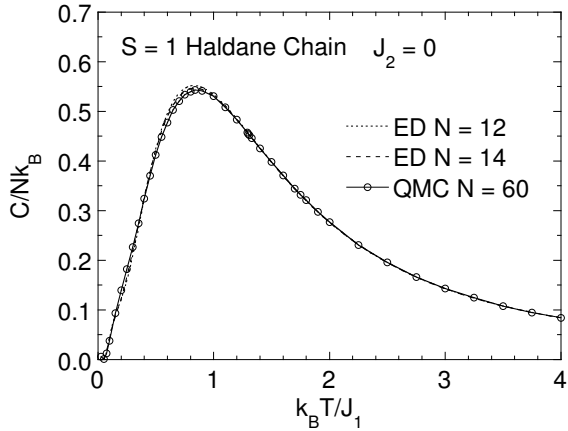}
\caption{Magnetic heat capacity $C$ versus temperature $T$ for the $S = 1$
$J_1$-$J_2$ chain with $J_2/J_1 = 0$ (``Haldane chain''), calculated using exact
diagonalization (ED) with $N = 10$ and 12, and quantum Monte Carlo (QMC)
simulations for $N = 60$.}
\label{FigJ1J2ChainJ2=0s}
\end{figure}

\begin{figure}[t]
\includegraphics[width=3in]{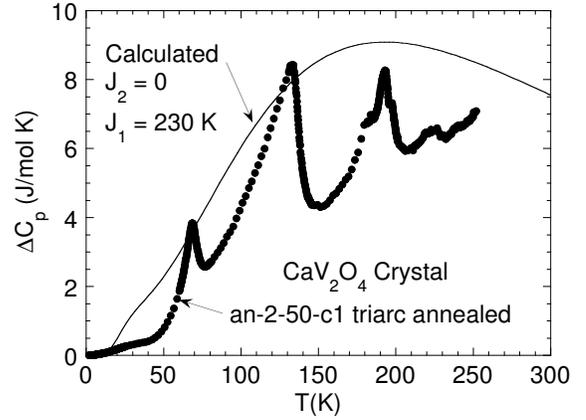}
\caption{Magnetic heat capacity $C$ versus temperature $T$ for the
spin $S = 1$ $J_1$-$J_2$ chain with $J_2/J_1 = 0$ and $J_1/k_{\rm B} = 230$~K,
calculated using exact diagonalization with $N = 14$ (solid curve) from
Fig.~\protect\ref{FigJ1J2ChainJ2=0s}.  The data points are the $\Delta C(T)$
data for annealed crystal an-2-50-c1 from
Fig.~\protect\ref{FigCaV2O4_DeltaCp_Shift_All}.}
\label{FigCaV2O4_N=14_DeltaC_Fit}
\end{figure}

The magnetic heat capacity $C$ versus temperature $T$ was calculated by exact
diagonalization for $N = 12$ spins $S = 1$ over the range $-1 \leq J_2/J_1 \leq 5$. 
Representative results are plotted in Fig.~\ref{FigJ1J2ChainN_12Cmag}.  The
variation in $C(T)$ with $N$ is illustrated in Fig.~\ref{FigJ1J2ChainJ2=0s} for
$J_2/J_1 =0$ and $N = 12$ and 14 from exact diagonalization calculations, and
for $N = 60$ from quantum Monte Carlo simulations.  The data for the different
$N$  are seen to be nearly the same.  The values of the maxima
$C^{\rm max}$ in the magnetic heat capacity and the temperatures $T_C^{\rm
max}$ at which they occur are listed above in
Table~\ref{TableJ2=0ChainChiCmax}.  Our results for the specific heat are 
consistent with previous transfer-matrix renormalization-group
computations.\cite{Xiang,FW05}  The two transfer-matrix
renormalization-group results differ at {\em high} temperatures.  Our QMC
results for $C$ obtained from rings with $N=60$ sites are in better agreement
with the older results which apply to the infinite $N$ limit\cite{Xiang} than
the more recent results obtained for open chains with $N=64$
sites.\cite{FW05}  The $C(T)$ data in Ref.~\onlinecite{FW05} were calculated
from a numerical derivative which resulted in systematic errors in the data
at high temperatures.\cite{FW08}

We cannot confidently derive the exchange constants in CaV$_2$O$_4$ from fits of
our heat capacity data by the theory.  Extraction of the magnetic part of the
experimental heat capacity is tenuous because of the presence of the
orthorhombic to monoclinic structural transition at $T_{\rm S}\approx 150$~K
and the transition(s) at $T_{\rm S1}\approx 200$~K\@.  Therefore
the relationship of the heat capacities $\Delta C(T)$ in
Figs.~\ref{DeltaCp_DeltaS_All}(a) and
\ref{FigCaV2O4_DeltaCp_Shift_All} to the magnetic heat capacities of the
samples is unclear.  Here, we will just compare the
theoretical magnetic heat capacity calculated for the exchange constants $J_1 =
230$~K and $J_2 = 0$, that were already deduced in the previous section, with
the experimental $\Delta C(T)$ data to see if theory and experiment are at least
roughly in agreement.  This comparison is shown in
Fig.~\ref{FigCaV2O4_N=14_DeltaC_Fit} for annealed CaV$_2$O$_4$ crystal
an-2-50-c1 from Fig.~\ref{FigCaV2O4_DeltaCp_Shift_All}.  Overall, the theory and
experimental data have roughly the same magnitude, but the data are
systematically below the theoretical prediction.  This is likely caused by the
heat capacity of the nonmagnetic reference compound CaSc$_2$O$_4$ being 
somewhat different from the lattice heat capacity of CaV$_2$O$_4$.   We note
from Fig.~\ref{Cp} that  a difference of 5~J/mol~K between the theory and
experiment in Fig.~\ref{FigCaV2O4_N=14_DeltaC_Fit} is only about 5\% of the
total heat capacity of the samples at 200~K\@.  In addition, we have not
included in the theory interchain couplings that lead to long-range
antiferromagnetic order, or the effect of the magnetic ordering on the heat
capacity including the effect of the energy gap in the spin wave spectrum below
$T_{\rm N}$.

\subsection{Interchain Coupling}

Within the $S = 1$ $J_1$-$J_2$ Heisenberg spin chain model, we found above that
$J_2/J_1 \approx 0$ and $J_1 \approx 230$~K in CaV$_2$O$_4$ near room temperature.  Thus the
crystallographic zigzag vanadium chains in CaV$_2$O$_4$ act like $S = 1$
linear spin chains with nearest-neighbor interaction $J_1$.  This is a
so-called Haldane chain\cite{Haldane} with a nonmagnetic singlet ground state
and an energy gap for spin excitations given by\cite{GJL94, White1993}
$\approx 0.4105 J_1$.  An interchain coupling $J_\perp$ must be present in
order to overcome this spin gap and induce long-range antiferromagnetic ordering
at $T_{\rm N}$.  Pedrini {\it et al.} \cite{Pedrini2004,Pedrini2007} have
recently estimated the dependence of $T_{\rm N}/J_1$ on $J_\perp/J_1$ using a
random-phase approximation for the interchain coupling for $S=1$ Haldane
chains.  Using our values $T_{\rm N} = 68$~K, $J_1 = 230$~K and $T^{\rm max} =
300$~K, we obtain $J_\perp/J_1 \approx 0.04$ and $J_\perp = 5$--10~K\@. However,
it should be emphasized that the treatment
of Refs.~\onlinecite{Pedrini2004} and~\onlinecite{Pedrini2007} assumes a
nonfrustrated interchain coupling geometry such that the result $J_\perp
\approx 10$~K should be considered as a lower bound.  Still, the  value of
$J_\perp/J_1$ is sufficiently small that a redetermination of $J_2/J_1$ and
$J_1$, from a $J_1$-$J_2$-$J_\perp$ model fitted to the observed susceptibility
data for CaV$_2$O$_4$ near room temperature, would yield very similar values of
$J_2/J_1$ and $J_1$ to those we have already estimated using the isolated chain
$J_1$-$J_2$ model.  Additional and more conclusive information about the interchain coupling strength(s) will become available from analysis of inelastic neutron scattering measurements of the magnetic excitation dispersion relations.\cite{Pieper2008}

\section{\label{summary}Summary} 

We have synthesized the $S = 1$ spin chain compound CaV$_2$O$_4$ in high purity
polycrystalline form and as single crystals.
Our magnetic susceptibility $\chi(T)$ and ac magnetic susceptibility $\chi_{\rm
ac}(T)$ measurements do not show any signature of a spin glass-like transition
around 20~K that was previously reported.\cite{fukushima, kikuchi}  We instead
observe long-range antiferromagnetic ordering at sample-dependent N\'eel
temperatures $T_{\rm N}\approx 50$--70~K as shown in Table~\ref{ordering_temp}. 
The N\'eel temperature and the orthorhombic-to-monoclinic structural transition temperature $T_{\rm S}$ in Table~\ref{ordering_temp} both show a large systematic
variation between different samples.  Those temperatures for an unannealed
crystal are each less than those for an annealed crystal which in turn are less
than those for a  sintered polycrystalline sample.  The cause of these large
temperature differences, especially between as-grown and annealed single
crystals, is unclear.  The transition temperature differences may arise from
small changes in oxygen stoichiometry ($\lesssim 1$ at.\%, below the threshold
of detection by TGA or XRD) and/or from structural strain, both of which may be
reduced upon annealing the as-grown crystals at 1200~$^\circ$C in 5\%~H$_2$/He. 
In addition, other small chemical differences and/or structural defects may be
relevant.

Our heat capacity $C_{\rm p}(T)$, linear thermal expansion $\alpha(T)$, and
$\chi(T)$ measurements reveal distinct features at the orthorhombic-to-monoclinic structural transition temperature $T_{\rm S}$ identified from
our diffraction studies.\cite{bella}  We inferred from a combination of structural studies and physical property measurements that the origin of the third transition at $T_{\rm S1} \approx 200$~K in one of our annealed crystals was mostly due to the structural transition in the V$_2$O$_3$ impurity phase that grew coherently upon annealing the crystal.  In the other annealed crystal, we ruled out this source and we are thus left with a transition at $T_{\rm S1}$ with unknown origin.  We speculate that this transition may be the long-sought chiral phase transition originally postulated by Villain in 1977.\cite{Villain1977} 

The $\chi(T)$ shows a broad maximum at about 300~K indicating short-range
antiferromagnetic (AF) ordering in a low-dimensional antiferromagnet as
previously observed\cite{fukushima, kikuchi} and the $\chi(T)$ above $T_{\rm
N}$ in single crystals is nearly isotropic.  The anisotropic $\chi(T)$ below
$T_{\rm N}$ shows that the (average) easy axis of the antiferromagnetic
structure is the orthorhombic $b$-axis.  The magnetic spin susceptibility along this axis is found to be a large finite value for $T\rightarrow 0$, instead of being zero as expected for a classical collinear antiferromagnet. 
This result is consistent with our observed noncollinear
magnetic structure below $T_{\rm N}$.\cite{zong,Pieper2008B}  In view of the fact that CaV$_2$O$_4$ is a low-dimensional spin system, quantum fluctuations could also contribute to both the observed reduced zero temperature ordered moment and the relatively large zero temperature spin susceptibility.

We analyzed the $\chi(T)$ data near room temperature in terms of theory for the $S =
1$ $J_1$-$J_2$ linear Heisenberg chain, where $J_1$($J_2$) is the
(next-)nearest-neighbor interaction along the chain.  We obtain $J_1/k_{\rm B}
\approx 230$~K, but surprisingly $J_2/J_1 \approx 0$ (or $J_1/J_2 \approx 0$), so the exchange
connectivity of the spin lattice appears to correspond to linear $S = 1$ Haldane
chains instead of zigzag spin chains as expected from the crystal structure.  This result is consistent with analysis of our high temperature (up to 1000~K) magnetic susceptibility measurements on single crystal CaV$_2$O$_4$.\cite{Pieper2008B}  We estimated here the coupling $J_\perp$ between these chains that leads to long-range
AF order at $T_{\rm N}$ to be $J_\perp/J_1 \gtrsim 0.04$, i.e., only
slightly larger than the value $J_\perp/J_1 \approx 0.02$
needed\cite{Pedrini2004, Pedrini2007} to eliminate the energy gap (Haldane gap)
for magnetic excitations.  

From our $C_{\rm p}(T)$ measurements, the estimated molar magnetic entropy at
$T_{\rm N}$ is only $\approx 8\%$ of its maximum value 2Rln($2S+1$) = 2Rln(3),
where R is the molar gas constant, and the heat capacity jump at
$T_{\rm N}$ is only a few percent of the value expected in mean field
theory for $S = 1$.  Both results indicate strong short range antiferromagnetic
order above $T_{\rm N}$ and large values $J_1$ and/or $J_2 > 100$ K, consistent
with the $\chi(T)$ data.  We also compared the $C_{\rm p}(T)$ data with the
theoretical prediction for the magnetic heat capacity using the exchange
constants found from the magnetic susceptibility analysis, and rough agreement
was found.  However, this comparison is not very precise or useful because the
structural transition at $T_{\rm S} \sim 150$~K and the transition(s) at $T_{\rm S1} \sim 200$~K for our two annealed single crystals, make large contributions to
$C_{\rm p}(T)$.  In addition, the accuracy of the measured heat capacity of
the nonmagnetic reference compound ${\rm CaSc_2O_4}$ in representing the lattice
heat capacity of ${\rm CaV_2O_4}$ is unknown.  Thus extracting the magnetic
part of the heat capacity at high temperatures from the observed $C_{\rm p}(T)$ data for comparison with theory is ambiguous.  

In closing, we note the following additional issues that could usefully be
addressed in future work.  Our analyses of our $\chi(T)$ data for CaV$_2$O$_4$
to obtain the exchange constants $J_1$ and $J_2$ were based on fitting the
experimental $\chi(T)$ data only near room temperature, since our calculations
of $\chi(T)$ all showed nonmagnetic singlet ground states, contrary to
observation, and could not reproduce the observed antiferromagnetic ordering at
low temperatures.  Calculations containing additional interactions (see also
below) and/or anisotropies are needed for comparison with the lower temperature
data.

The orthorhombic crystal structure of CaV$_2$O$_4$ at room temperature contains
two crystallographically inequivalent but similar  V$^{+3}$ $S = 1$ zigzag
chains.   These chains may therefore have different exchange constants
associated with each of them.  For simplicity, our $\chi(T)$ data were analyzed
assuming a single type of zigzag chain.  Furthermore, the extent to which the
transitions at $T_{\rm S1}$ and $T_{\rm S}$ affect the magnetic interactions is
not yet clear.  

From crystal structure considerations, one expects that $J_2/J_1 \approx
1$ in CaV$_2$O$_4$,\cite{fukushima, kikuchi} instead of $J_2/J_1 \approx 0$ as
found here.  This suggests that additional magnetic interactions and/or
anisotropy terms beyond the Heisenberg interactions $J_1$ and $J_2$ and
interchain coupling $J_\perp$ considered here may be important.  In addition to 
single ion anisotropy and other types of anisotropy, we mention as possibilities
the Dzyaloshinskii-Moriya interaction, biquadratic exchange, and 
cyclic exchange interactions within the zigzag chains.  When such
additional terms are included in the analysis, the fitted value of $J_2/J_1$
could turn out to be closer to unity.  A four-spin cyclic exchange interaction
has been found to be important to the magnetic susceptibility in cuprate spin
ladders.\cite{Johnston2000}  In these spin ladders, there are 
exchange interactions $J$ and $J^\prime$ between nearest-neighbor Cu$^{+2}$
spins~1/2 along the legs and across the rungs of the spin ladder,
respectively.  For the $S = 1/2$ two-leg ladder compound ${\rm SrCu_2O_3}$,
if only $J$ and $J^\prime$ are included in fits to the data, one obtains
$J^\prime/J \approx 0.5$.\cite{Johnston2000}  However, by also including the
theoretically derived cyclic four-spin exchange interaction, the ratio
$J^\prime/J$  increases from 0.5 to a value closer to unity, as expected from
the crystal structure.

Pieper \emph{et al.}\ have recently proposed a very different and very interesting model to explain the inference  that $J_2/J_1 \approx 0$ around room temperature which involves partial orbital ordering of the two $d$-electrons of V among the three $t_{2g}$ orbitals.\cite{Pieper2008B}  Furthermore, in order to explain the magnetic structure at low temperatures, they deduce that the nature of the orbital ordering changes below $T_{\rm S}$ such that the effective spin lattice becomes a spin-1 two-leg ladder.

It has been well documented that fits of magnetic susceptibility
data by theory tests only the consistency of a spin model with the
data, and not the uniqueness of the model.  A good example of this fact arose
in the study of the antiferromagnetic alternating exchange chain compound
vanadyl pryophosphate, ${\rm (VO)_2P_2O_2}$, the history of which is described
in detail in the introduction of Ref.~\onlinecite{Johnston2001}.  The ultimate
arbiter of the validity of a spin model is inelastic neutron scattering
measurements of the magnetic excitation dispersion relations in single
crystals.  Theoretical
calculations of the exchange interactions are much needed and would also be
valuable in this regard.

Finally, the origin of the intrinsic heat capacity anomalies at $T_{\rm S1} \approx 200$~K for the two \emph{annealed} single crystals of CaV$_2$O$_4$ needs to be further studied.  We speculate that this transition may be the long-sought chiral phase transition originally postulated by Villain in 1977.\cite{Villain1977}

Note added---After this work and this paper were nearly completed, Sakurai reported a very interesting and detailed study of the magnetic and electronic phase diagram of polycrystalline samples of the solid solution Ca$_{1-x}$Na$_x$V$_2$O$_4$ prepared under high pressure.\cite{Sakurai2008}

\begin{acknowledgments}  

We acknowledge useful discussions with R. McQueeney, and we thank D. Robinson for the excellent technical support of our high-energy x-ray diffraction study.  Work at Ames Laboratory was supported by the United States Department of Energy-Basic Energy Sciences under Contract No.~DE-AC02-07CH11358.  Use of the Advanced Photon Source (APS) was supported by the U.S. Department of Energy, Office of Science, under Contract No.~DE-AC02-06CH11357. The Midwest Universities Collaborative Access Team (MUCAT) sector at the APS is supported by the U.S. Department of Energy, Office of Science, through the Ames Laboratory under Contract No. DE-AC02-07CH11358.  The work of A.H. was supported by the Deutsche Forschungsgemeinschaft through a Heisenberg fellowship and grant No.~HO 2325/4-1.  M.R. acknowledges fundings from Deutsche Forschungsgemeinschaft (grant UL 164/4).

\end{acknowledgments}

\end{document}